\documentclass[12pt]{article}
\usepackage{epsfig,epsf}
\begin{document}
\def\beq{\begin{equation}}
\def\eeq{\end{equation}}
\def\bea{\begin{eqnarray}}
\def\eea{\end{eqnarray}}
\def\eq#1{{Eq.~(\ref{#1})}}
\def\fig#1{{Fig.~\ref{#1}}}
\newcommand{\bas}{\bar{\alpha}_S}
\newcommand{\as}{\alpha_S} 
\newcommand{\bra}[1]{\langle #1 |}
\newcommand{\ket}[1]{|#1\rangle}
\newcommand{\bracket}[2]{\langle #1|#2\rangle}
\newcommand{\intp}[1]{\int \frac{d^4 #1}{(2\pi)^4}}
\newcommand{\mn}{{\mu\nu}}
\newcommand{\tr}{{\rm tr}}
\newcommand{\Tr}{{\rm Tr}}
\newcommand{\T} {\mbox{T}}
\newcommand{\braket}[2]{\langle #1|#2\rangle}
\newcommand{\ab}{\bar{\alpha}_S}

\setcounter{secnumdepth}{7}
\setcounter{tocdepth}{7}
\parskip=\itemsep               %?

\setlength{\itemsep}{0pt}       %?
\setlength{\partopsep}{0pt}     %?
\setlength{\topsep}{0pt}        %?
%---layout fuer eine dina4 seite-------------------
\setlength{\textheight}{22cm}
\setlength{\textwidth}{174mm}
\setlength{\topmargin}{-1.5cm}

%\input psfig
%%%%%%%%%%%%%%%%%%%%%%%%%%%%%%%%%%%%%%%%%%%%%%%%%%%%%%%%%%%%%%%
%\renewcommand{\thefootnote}{\fnsymbol{footnote}}
\newcommand{\beqar}[1]{\begin{eqnarray}\label{#1}}
\newcommand{\eeqar}{\end{eqnarray}}
\newcommand{\m}{\marginpar{*}}
\newcommand{\lash}[1]{\not\! #1 \,}
\newcommand{\nn}{\nonumber}
\newcommand{\D}{\partial}
\newcommand{\h}{\frac{1}{2}}
\newcommand{\g}{{\rm g}}
\newcommand{\el}{{\cal L}}
\newcommand{\A}{{\cal A}}
\newcommand{\Ka}{{\cal K}}
\newcommand{\al}{\alpha}
\newcommand{\be}{\beta}
\newcommand{\ep}{\varepsilon}
\newcommand{\ga}{\gamma}
\newcommand{\de}{\delta}
\newcommand{\De}{\Delta}
\newcommand{\et}{\eta}
\newcommand{\ka}{\vec{\kappa}}
\newcommand{\la}{\lambda}
\newcommand{\ph}{\varphi}
\newcommand{\si}{\sigma}
\newcommand{\ro}{\varrho}
\newcommand{\Ga}{\Gamma} 
\newcommand{\om}{\omega}
\newcommand{\La}{\Lambda}  
\newcommand{\tG}{\tilde{G}}
\renewcommand{\theequation}{\thesection.\arabic{equation}}

%%%%%%%%%%%%%%%%%%%%%%%%%%%%%%%%%%%%%%%%%%%%%%%%%%%%%%%%%%%%%%%%%
% ABBREVIATED JOURNAL NAMES  
%
\def\ap#1#2#3{     {\it Ann. Phys. (NY) }{\bf #1} (19#2) #3}
\def\arnps#1#2#3{  {\it Ann. Rev. Nucl. Part. Sci. }{\bf #1} (19#2) #3}
\def\npb#1#2#3{    {\it Nucl. Phys. }{\bf B#1} (19#2) #3}
\def\plb#1#2#3{    {\it Phys. Lett. }{\bf B#1} (19#2) #3}
\def\prd#1#2#3{    {\it Phys. Rev. }{\bf D#1} (19#2) #3}
\def\prep#1#2#3{   {\it Phys. Rep. }{\bf #1} (19#2) #3}
\def\prl#1#2#3{    {\it Phys. Rev. Lett. }{\bf #1} (19#2) #3}
\def\ptp#1#2#3{    {\it Prog. Theor. Phys. }{\bf #1} (19#2) #3}
\def\rmp#1#2#3{    {\it Rev. Mod. Phys. }{\bf #1} (19#2) #3}
\def\zpc#1#2#3{    {\it Z. Phys. }{\bf C#1} (19#2) #3}
\def\mpla#1#2#3{   {\it Mod. Phys. Lett. }{\bf A#1} (19#2) #3}
\def\nc#1#2#3{     {\it Nuovo Cim. }{\bf #1} (19#2) #3}
\def\yf#1#2#3{     {\it Yad. Fiz. }{\bf #1} (19#2) #3}
\def\sjnp#1#2#3{   {\it Sov. J. Nucl. Phys. }{\bf #1} (19#2) #3}
\def\jetp#1#2#3{   {\it Sov. Phys. }{JETP }{\bf #1} (19#2) #3}
\def\jetpl#1#2#3{  {\it JETP Lett. }{\bf #1} (19#2) #3}
%%%%%%%%% notice the parenthesys is only on one side
\def\ppsjnp#1#2#3{ {\it (Sov. J. Nucl. Phys. }{\bf #1} (19#2) #3}
\def\ppjetp#1#2#3{ {\it (Sov. Phys. JETP }{\bf #1} (19#2) #3}
\def\ppjetpl#1#2#3{{\it (JETP Lett. }{\bf #1} (19#2) #3} 
\def\zetf#1#2#3{   {\it Zh. ETF }{\bf #1}(19#2) #3}
\def\cmp#1#2#3{    {\it Comm. Math. Phys. }{\bf #1} (19#2) #3}
\def\cpc#1#2#3{    {\it Comp. Phys. Commun. }{\bf #1} (19#2) #3}
\def\dis#1#2{      {\it Dissertation, }{\sf #1 } 19#2}
\def\dip#1#2#3{    {\it Diplomarbeit, }{\sf #1 #2} 19#3 }
\def\ib#1#2#3{     {\it ibid. }{\bf #1} (19#2) #3}
\def\jpg#1#2#3{        {\it J. Phys}. {\bf G#1}#2#3}  
%

%%%%%%%%%%%%%%%%%%%%%%%%%%%%%%%%%%%%%%%%%%%%%%%%%%%%%%%%%%%%%%%%%%%%%
%
%\renewcommand{\thefigure}{{\protect\bf\arabic{figure}}}
\def\thefootnote{\fnsymbol{footnote}} 
%
%
%
%\begin{titlepage}
\noindent
\begin{flushright}
\parbox[t]{10em}{ \tt TAUP \,\,2762/2004\\
 \today }
\end{flushright}
\vspace{1cm}
\begin{center}
{\LARGE  \bf The  Iancu-Mueller  factorization and}\\
~ \\
{\LARGE  \bf  high energy asymptotic behaviour}
\\

\vskip1cm {\large \bf ~M. ~Kozlov $ {}^{\dagger}$ \footnotetext{${}^{\ddagger}$ \,\,Email:
kozlov@tau.ac.il} and ~E. ~Levin ${}^{\,\star}$ \footnotetext{${}^{\,\star}$ \,\,Email:
leving@tau.ac.il, levin@mail.desy.de }} \vskip1cm

{\it HEP Department}\\
{\it School of Physics and Astronomy}\\
{\it Raymond and Beverly Sackler Faculty of Exact Science}\\
{\it Tel Aviv University, Tel Aviv, 69978, Israel}\\
\vskip0.3cm

\end{center}  
\bigskip
\begin{abstract} 	
We  show that the  Iancu-Mueller factorization has a simple interpretation 
in the Reggeon - like technique based on the BFKL Pomeron. The formula for 
calculating  the high energy asymptotic behaviour for the colour 
dipole-dipole amplitude is proposed  which suggests  a procedure to 
calculate this amplitude through the solution to the Balitsky-Kovchegov 
non-linear equation. We confirm the Iancu - Mueller result that a 
specific set of enhanced diagrams is responsible for the high energy 
behaviour  for  fixed QCD coupling. 
However, it is argued that  in the case of running QCD coupling, this 
asymptotic behaviour  originates  from the Balitsky-Kovchegov non-linear 
equation. A new solution to the non-linear equation is found which leads to a 
different asymptotic 
behaviour of the scattering amplitude even for  fixed $\as$. 

\end{abstract}

%*********************************************************************************  
%*********************************************************************************  
\newpage 

\def\thefootnote{\arabic{footnote}} 
\section{Introduction}
\label{sec:Introduction}
Iancu and Mueller in recent papers \cite{IM} have  suggested a new approach\footnote{The key 
ingredients of 
this approach have  been suggested by Mueller and Salam\cite{MS} but at that time the approach based 
mostly on numerical simulations while now we are able to develop  analytical methods.}  to determine  
the high 
energy 
asymptotic behaviour of the scattering amplitude which goes beyond  the non-linear equation (BK 
equation \cite{B,K})  of  the dense partonic system. In this system the gluon occupation numbers are 
large,
 the gluonic 
fields are strong and such a system enters a new phase of QCD: the colour glass condensate 
\cite{GLR,MUQI,MV,CGC} (see also Ref. \cite{IV} for a recent review of this approach). 
The key new element in Iancu-Mueller approach is an attempt to take into account  fluctuations in the 
partonic wave function of the fast moving particle which were neglected in the non-linear equation.
These fluctuations could be taken into account by more general approach to CGC related to so called JIMWLK 
\cite{CGC} equation, which is functional equation and it is not very practical at the moment. We believe 
that the Iancu-Mueller approach gives as an opportunity to find more transparent and analytic way to take 
into account the fluctuations in the partonic wave function.

The main goal of this paper is to show that the Iancu-Mueller approach arises naturally from the Reggeon 
- like diagram technique \cite{GLR}  based on the BFKL Pomerons \cite{BFKL}. This 
technique is the well known way of  incorporating  the fluctuation in the partronic wave function for 
high 
energy scattering  which leads to the non-linear BK  evolution equation \cite{BRAUN}. We will review the 
region of applicability of the non-linear equation for the scattering processes. It was discussed in 
Ref. \cite{GLR} (see also Refs.\cite{MR,BKL,LL})  but it has not  been utilized since that time.
 We show 
that the Iancu-Mueller result  for the scattering amplitude is valid  for frozen 
QCD coupling in a  large but limited range of energies, while for  running QCD coupling we expect 
rather the answer obtained in the 
colour glass condensate.

\begin{figure} [htb]
\begin{tabular}{l c  l}
\epsfig{file= 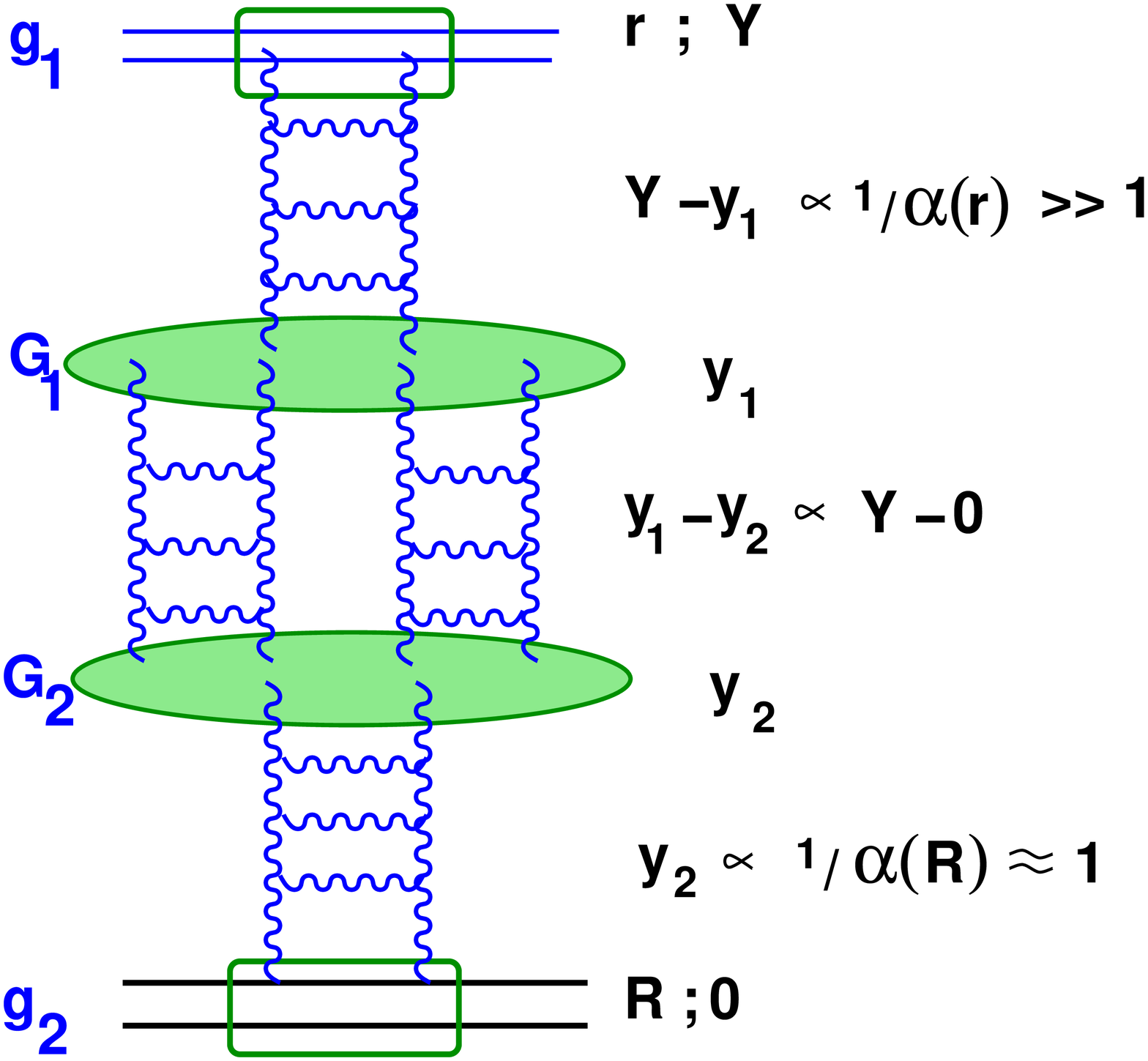,height=65mm}& \,\,\,\,\,\,\,\,\,\,\,\,\,\,\,\,\,
\,\,\,\,\,\,\,\,\,\,\,\,\,\,\,\,\,\,\,\,\,&
\epsfig{file= 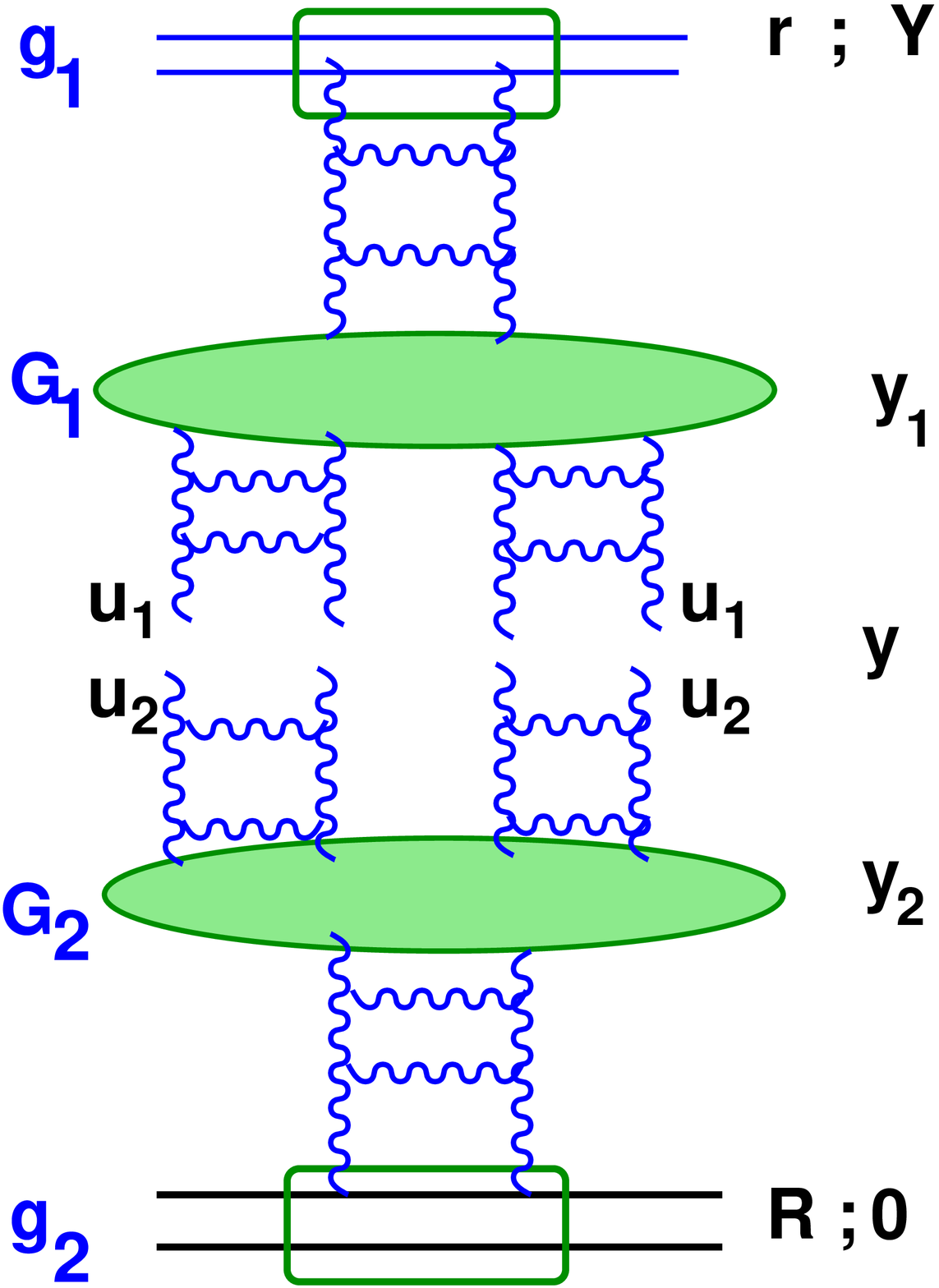,height=65mm} \\
       &  &  \\
\fig{enhdi}-a & & \fig{enhdi}-b \\
\end{tabular}
\caption{ The first enhanced diagram (\fig{enhdi}-a)  for dipole-dipole scattering and the Iancu-Mueller 
factorization for this diagram (\fig{enhdi}-b)}
\label{enhdi}
\end{figure}

We concentrate our efforts on understanding  dipole-dipole scattering.
The main idea of the Iancu-Mueller factorization as well as our approach to this factorization can 
be 
illustrated, considering the first so called enhanced diagram (see \fig{enhdi}-a) which describes the 
fluctuation in the partonic wave function of the incoming fast dipole.

We  calculate this diagram in the simple toy - model for the BFKL ladder, namely, assuming for 
the 
BFKL Pomeron with the intercept $\Delta$,  a  simple expression $e^{\Delta y}$, while  neglecting any 
dependence on the
 size of the interacting dipoles \cite{MUUN}. In addition, we impose the relation $ G_1 = g_1 = 
\Delta$ and 
$ G_2 = g_2 =\frac{\Delta}{N^2_c}$ where $N_c$ is the number of colours. These relations follow directly 
from the 
QCD estimates \cite{MUUN,LE},  but the most important $N_c$ suppression is the old result of topological 
expansion \cite{VE}.   

The expression for \fig{enhdi}-a has the form:
\bea \label{FENDISM}
A( \,\fig{enhdi}-a\,)\,\,&=&\,\,(-1)\,g_1\,g_2\,G_1\,G_2\,e^{\Delta Y}\,\left(\,\frac{1}{\Delta^2} 
\,\{\,e^{\Delta Y}\,\,-\,\,1\}\,\,-\,\,\frac{Y}{\Delta}\,\,\right)\,\,\nonumber \\
 &=&  -  \,\, \frac{\Delta^2}{N^4_c}\,e^{\Delta Y}\,\left(\,\{\,e^{\Delta
Y}\,\,-\,\,1\}\,\,-\,\,\Delta\,Y\,\,\,\right) 
 \eea

 In \eq{FENDISM} the sign minus reflects
the fact that such diagrams describe shadowing which   tames the power-like energy increase 
$e^{\Delta Y}$,   caused by
the BFKL Pomeron exchange.  For large $N_c$ there exists a region of energy given by the
inequality
 \beq \label{RENC} 
N^2_c \,\,\,\gg\,\,\, \frac{1}{N^2_c}e^{\,\Delta Y} \,\,\,\geq\,\,1
 \eeq
 in
which only the first term in \eq{FENDISM} is large (of the order of unity),  while all other terms are 
small.
In this region of energy the diagram has a simple form 
\beq \label{FENDISM1}
 A(\,\fig{enhdi}-a\,)\,\,=\,\,(-1)\, \frac{\Delta^2}{N^4_c}\,e^{2\,\Delta Y}
\eeq

The main idea of Iancu and Mueller \cite{IM} is to calculate the amplitude of \eq{FENDISM1} using a 
different approach, which is presented graphically in \fig{enhdi}-b. This approach is based on the
 generating 
function  \cite{MUUN,LE,LELA}
\beq   \label{Z}
Z(y,u)\,\,=\,\,\sum_n\,\,P_n(Y - y)\,u^n
\eeq
where $P_n(y)$ is the probability of  having  $n$-dipoles inside  the fast dipole.  We will show in 
section
3 that in the simple model, in which we have calculated  the diagram of \fig{enhdi}-a, the term in 
the 
generating function which is responsible for the amplitude for  production of two dipoles (see 
\fig{enhdi}-b) 
has the following form:
\beq \label{AMSM}
A(Y-y,\gamma)\,\,=\,\,1 - Z(Y - y,u\, \equiv \, 1 + \gamma) \,\,=\,\,(-1)\,\gamma^2 \,e^{\Delta (Y - 
y)}\,\left( 
e^{\Delta (Y - y)}\,\,-\,\,1\right)
\eeq
Therefore, \eq{FENDISM1} can be written in a different form, namely
\bea \label{IMSM} 
A(\,\fig{enhdi}-a\,)&=&\Delta^2\,\,(-1)\,A(Y-y,\gamma=1)\,A(y,\gamma= \frac{1}{N^2_c}) \nonumber \\
 &=&\Delta^2\,(-1) P_2(Y-y)\,P_2(y)\,\frac{1}{N^4_c}\,\\
 &=&\frac{(-1)\,\Delta^2}{4}\, \frac{d^2 \left(1 - Z(Y - y, 1 + \gamma) 
\right)}{(d \gamma)^2}|_{\gamma=0}\,\,\frac{d^2 \left(1 - Z(y, 1 + \gamma/N^2_c) \right)}{(d \gamma)^2}
|_{\gamma=0} \nonumber
\eea
\eq{IMSM} leads to the answer, which is not the same as the correct expression for the 
enhanced diagram given 
in \eq{FENDISM}, but  it correctly  reproduces    the leading term 
(see \eq{FENDISM1} ). The corrections to this term 
depend on the choice of the value for the rapidity $y$ (see \fig{enhdi}-b) which has only  auxiliary meaning.
The fact that the  generating function $Z$ depends on $y$ is very transparent,  since the wave function 
of 
the 
dipole depends on the reference frame. However, the amplitude as a physical observable should not depend 
on 
$y$. It does not depend on $y$ provided
\beq \label{ACSM}
\frac{1}{N^4_c} \,e^{\Delta(Y - 
y)}\,\,\approx 
\frac{1}{N^2_c}\,\,\ll\,\,1\,\,\,\,\,\mbox{and}\,\,\,\,\,\,\,\frac{1}{N^4_c} 
\,e^{\Delta( y)}\,\,\approx
\frac{1}{N^2_c}\,\,\ll\,\,1
 \eeq
The choice $y = Y/2$ leads to the estimates that both terms in  \eq{ACSM} 
are 
 of the order of $1/N^2_c \,\ll\,1$,  and  is most accurate.  This simple seems to
 justify  the choice of the reference frame suggested in Ref. \cite{IM}.

Formula of \eq{IMSM} is the simplest example of  the Iancu-Mueller factorization. In section 3 we 
will 
consider in detail a toy-model which we have used here to illustrate  the main idea of the 
Iancu-Mueller 
factorization. We will show that \eq{IMSM} has a natural generalization which coincides with the direct 
sum of 
all enhanced diagrams in the kinematic region given by \eq{RENC}. In section 4 the model result will be 
generalized 
for 
the QCD case to include  a dependence on the sizes of the interacting  dipoles. 

\section{Non-linear equation}
\begin{figure} [htb]
\begin{tabular}{l l}
\epsfig{file= 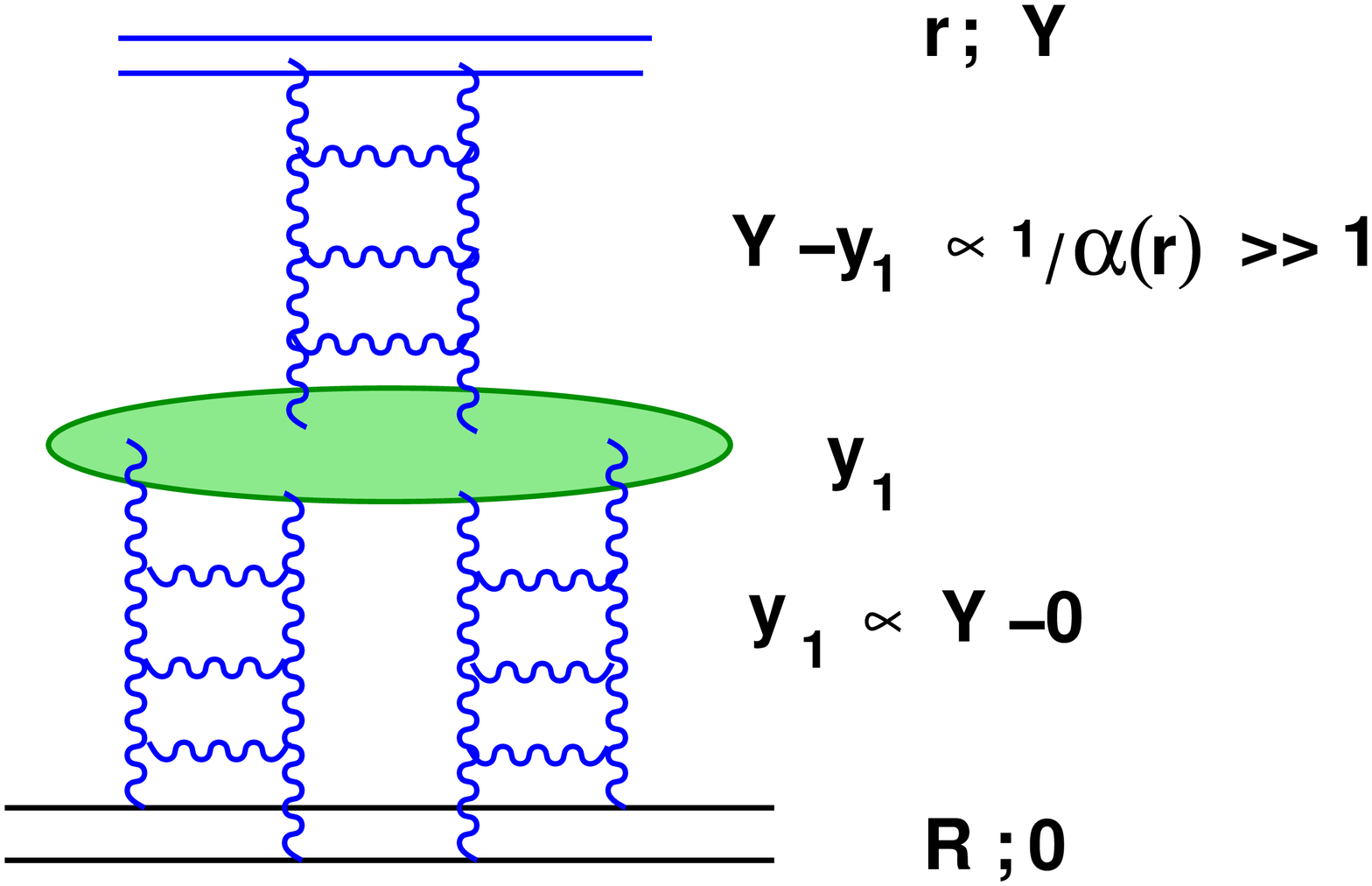,height=55mm}&
\epsfig{file= 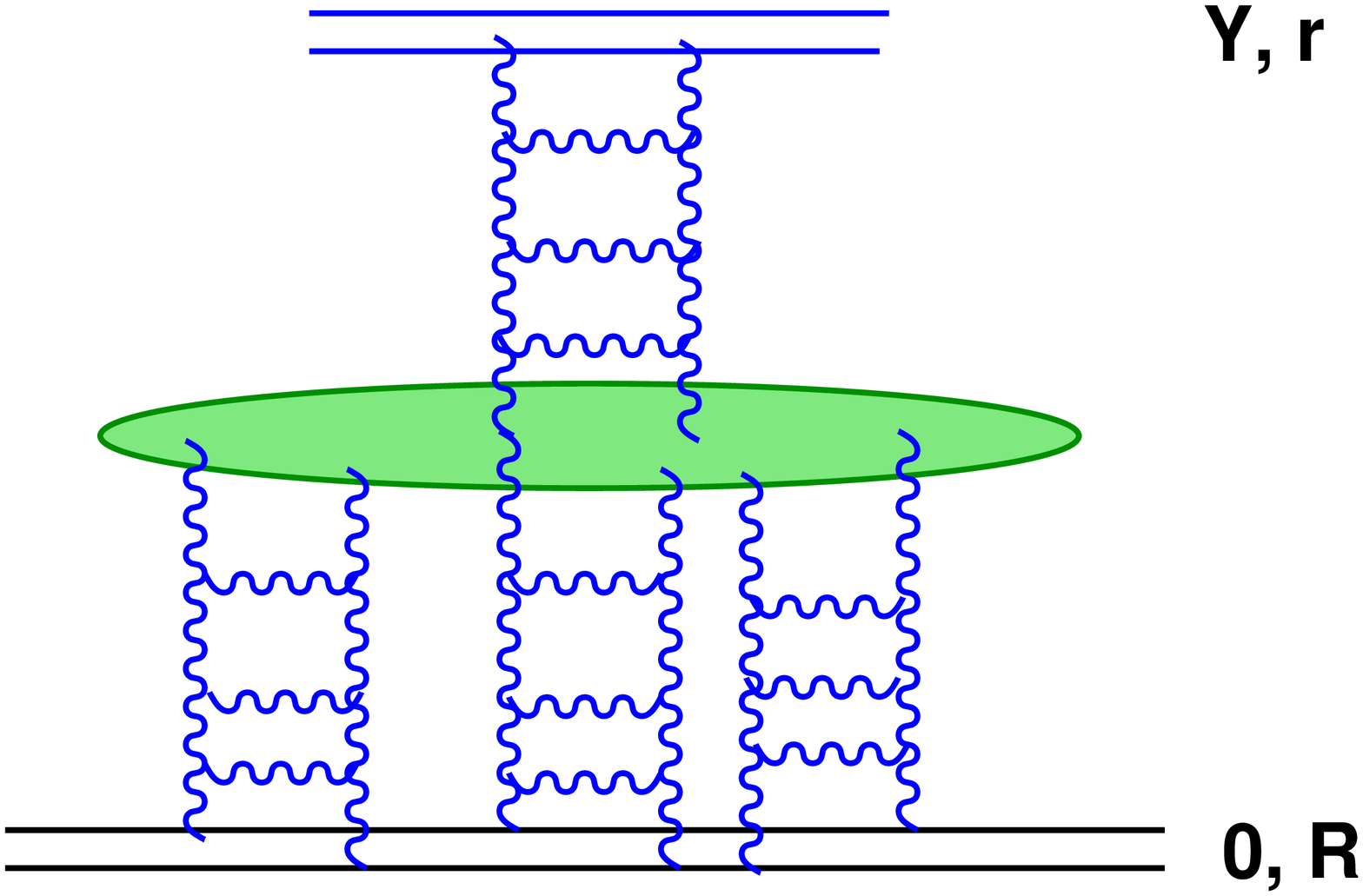,height=55mm} \\
       &    \\
\fig{fandi}-a & \fig{fandi}-b \\
\end{tabular}
\caption{ The first `fan'  diagram (\fig{fandi}-a)  for dipole-dipole scattering that is summed by 
the non-linear evolution equation,  and the first diagram that is not included in summation by the 
non-linear equation (\fig{fandi}-b)}
\label{fandi}
\end{figure}

The non-linear equation \cite{GLR,MUQI}, the final form of which at fixed impact parameter  was 
suggested by Balitsky \cite{B} and Kovchegov \cite{K},  sums all  so called `fan' diagrams (see 
\fig{fandi}-a). We would like to repeat here  the arguments that led to this equation  \cite{GLR,MUQI} 
for the case of the dipole-dipole scattering. The derivation given in 
Refs. \cite{B,K}  used the fact that the target is a heavy nucleus. As it will be 
clarified 
later, 
the nuclear target indeed enlarges the region of applicability for the non-linear equation, but  it is 
not 
essential for the saturation in  the colour glass condensate domain.

\subsection{The first enhanced diagram:}
\begin{figure}
\begin{center}
\epsfig{file=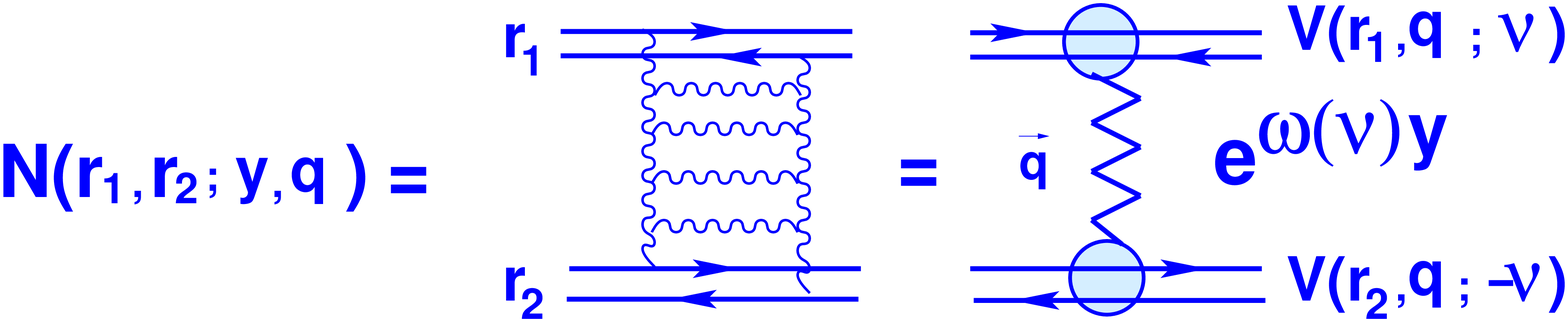,width=160mm}
\end{center}
\caption{The BFKL Pomeron.}
\label{ops}
\end{figure}

For a deeper  understanding of the non-linear equation we  start with the  calculation of  the same 
enhanced diagram  of \fig{enhdi}-a. The gluon `ladder' shown in this figure is the solution to the 
BFKL equation \cite{BFKL}, which can be written for the dipole ($r_1$)-dipole($r_2$)  amplitude  in the 
factorized form \cite{BRAUN,LIP,BFLLRW,NP}
 (see
\fig{ops} ).
\beq \label{BFKL1}
N^{BFKL}( r_{1, t},r_{2, t}; y, q)\,\,= \,\,\frac{\as^2}{4}\,\,
\eeq
$$
\,\,\int \frac{d \nu}{2\,\pi\,i}\,\,D(\nu)\,\, 
e^{\omega(\nu)\,y}
\,V(r_{1,t}, q ;\nu)\,V(r_{2,t}, q ;-\nu)
$$
with
\beq \label{D}
D(\nu)\,\,\,=\,\,\,\frac{\nu^2}{ (\nu^2 + \frac{1}{4})^2}
\eeq
 and with
\beq \label{OMEGA}
\omega(\nu)\,\,\equiv\,\,\frac{\as N_c}{\pi} \,\chi(\gamma)\,\,=\,\, 
\frac{\as N_c}{\pi}\,\left(\,2 
\,\psi(1) \,-\,\psi(\gamma)\,-
\,\psi(1 - \gamma)\,\right)\,;
\eeq
 where $\psi(f) \,=\,d \ln \Gamma(f)/d f$,  $ \Gamma(f)$ is the Euler gamma function, 
$\gamma\,=\,\h\,-\,i\,\nu$  and where
\beq \label{V}
V(r_{i,t}, q ;\nu)\,\,=\,\,\frac{2 \pi^2}{r_{i,t}\,\,b(\nu)}\,\,\int d^2 R \,e^{i 
\vec{q} \cdot \vec{R}}\,\,\left( 
\,\frac{r^2_{i,t}}{(\vec{R}_i
\,+\,\frac{1}{2}\vec{r}_{i,t})^2\,\,
(\vec{R}_i\,-\,\frac{1}{2}\vec{r}_{i,t})^2}\,\right)^{\frac{1}{2}
\,-\,i\,\nu}
\eeq
and   the following is our  notation: $y\,=\,\ln(x_0/x)$;\,\,$r_{i.t}$  is the size of the colour
dipole $``i"$ and $R_i$ is the position of the center of mass of  this dipole. $q$ is momentum 
transferred 
along the BFKL Pomeron (see \fig{ops} ).

The coefficient $b(\nu)$ in \eq{V} is defined in Refs. \cite{LIP,NP},  but we do not need  its 
explicit form in what we discuss 
below. The only important property of \eq{V} that we will use is \cite{LIP,NP}
\beq \label{VV}
\frac{1}{(2 \pi)^2}\,\,\int\,\frac{d^2 r_{t}}{r^2_t}\,\,V\left(r_t,q; \nu 
\right)\,,V\left(r_t,q; - \nu, 
\right)\,\,\,=\,\,\,\delta (\nu - \nu')
\eeq

The simplest enhanced diagram of \fig{bfklenh} has been given in Ref. \cite{NP} and it is equal to
\beq \label{BFKLENH}
N^{enh}\left(r_{1,t},r_{2,t};y,q \right)\,\,=\,\,-\,\frac{\as^4\,\pi^4}{8}\,\left(\frac{\as N_c}{2 
\pi^2}\right)^2 \,\,\int\,d^2 k\,\, d \gamma \,d \gamma_1 \,\,d \gamma_2\,\int^{Y} \,d y_1 
\,\int^{y_1}\,d y_2 \,\,
\eeq
$$
V(r_{1,t},q;\nu )
 D(\nu)\,
e^{ \omega(\nu)\,(Y  - y_1)}
\,G_{3P}( \nu; \nu_1,\nu_2; q,k )\, D(\nu_1)\,D(\nu_2)\,e^{ (\omega(\nu_1)\,+\,\omega(\nu_2))(y_1  - 
y_2)}G_{3P}( \nu; \nu_1,\nu_2; q,k )
$$
$$
\,D(\nu)\,e^{ \omega(\nu)\,(y_2)}\,V(r_{1,t},q;\nu )
$$
where $\gamma_i = 1/2 + i \nu_i$ and the triple Pomeron vertex  $G_{3P}$ is calculated in Ref. \cite{NP} 
 all details about \eq{BFKLENH} as well as its explicit derivation,  is given.

In spite of the additional  complicated integrations over momenta transferred and 
anomalous 
dimensions $\gamma_i$, appearing  in  \eq{BFKLENH}, the equation    has the same integrations over 
rapidities as in the simple 
toy model. Calculating the integrals over $y_1$ and $y_2$  and 
 keeping only the maximal power of energy,  we reduce \eq{BFKLENH} to a simpler expression
\beq \label{BFKLENH1}
N^{enh}\left(r_{1,t},r_{2,t};y,q \right)\,\,=\,\,-\,\frac{\bar{\alpha}^4_S\,}{32\,N^4_c}\,\int\,d^2 
k\,\, d 
\gamma \,d \gamma_1 \,d \gamma_2\,\int^Y \,dy_1
\,\int^{y_1}\,d y_2 \,\,V(r_{1,t},q;\nu )
\eeq
$$
 D(\nu)\,
\,G_{3P}( \nu; \nu_1,\nu_2; q,k )\, D(\nu_1)\,D(\nu_2)\,\frac{1}{ (\chi(\nu) - \chi(\nu_1) - 
\chi(\nu_2))^2}\,  e^{ (\omega(\nu_1)\,+\,\omega(\nu_2))\, Y 
}G_{3P}( \nu; \nu_1,\nu_2; q,k )
$$
$$
\,D(\nu)\,\,V(r_{1,t},q;\nu )
$$
Comparing \eq{BFKLENH1} with \eq{BFKL1} one can see that the ratio of these diagrams is of the order of
\beq \label{RAT}
\frac{N(\fig{bfklenh})}{N(\fig{ops})}\,\,\propto\,\,\frac{\bar{\alpha}^2_S}{N^2_c}\,e^{ 
\bar{\alpha}_S\,\chi(0)\,Y}  
 \eeq
At large values of $Y$ this ratio becomes of the order of unity,  and to obtain a correct scattering 
amplitude the enhanced diagrams should be 
taken into account.

\subsection{`Fan' diagrams:}

At first sight this  discussion leaves no room  for a special status of the   `fan' diagrams 
(see 
\fig{fandi} ) that are  summed by the non-linear equation \cite{B,K}. Indeed, as  was analyzed
long ago \cite{GLR},  we cannot 
expect any suppression of the enhanced diagrams for the fixed QCD coupling,  which has been used in the 
above calculations. However, the situation changes crucially if we consider 
the scattering amplitude for two dipoles with  different sizes (say $r_{1,t}\equiv r \,\,\ll\,\,r_{2,t} 
\equiv R $ ) and we 
 take into account  a  running QCD coupling. For rather rough estimates let us assume that we can 
include  running $\as$ by replacing \eq{OMEGA} by 
\beq \label{OMEGARUN}
\omega(\nu )\,\,=\,\,\as(r)\,\chi(\nu )\,\,.
\eeq

\begin{figure}[htb]
\begin{minipage}{10cm}
\begin{center}
\epsfig{file=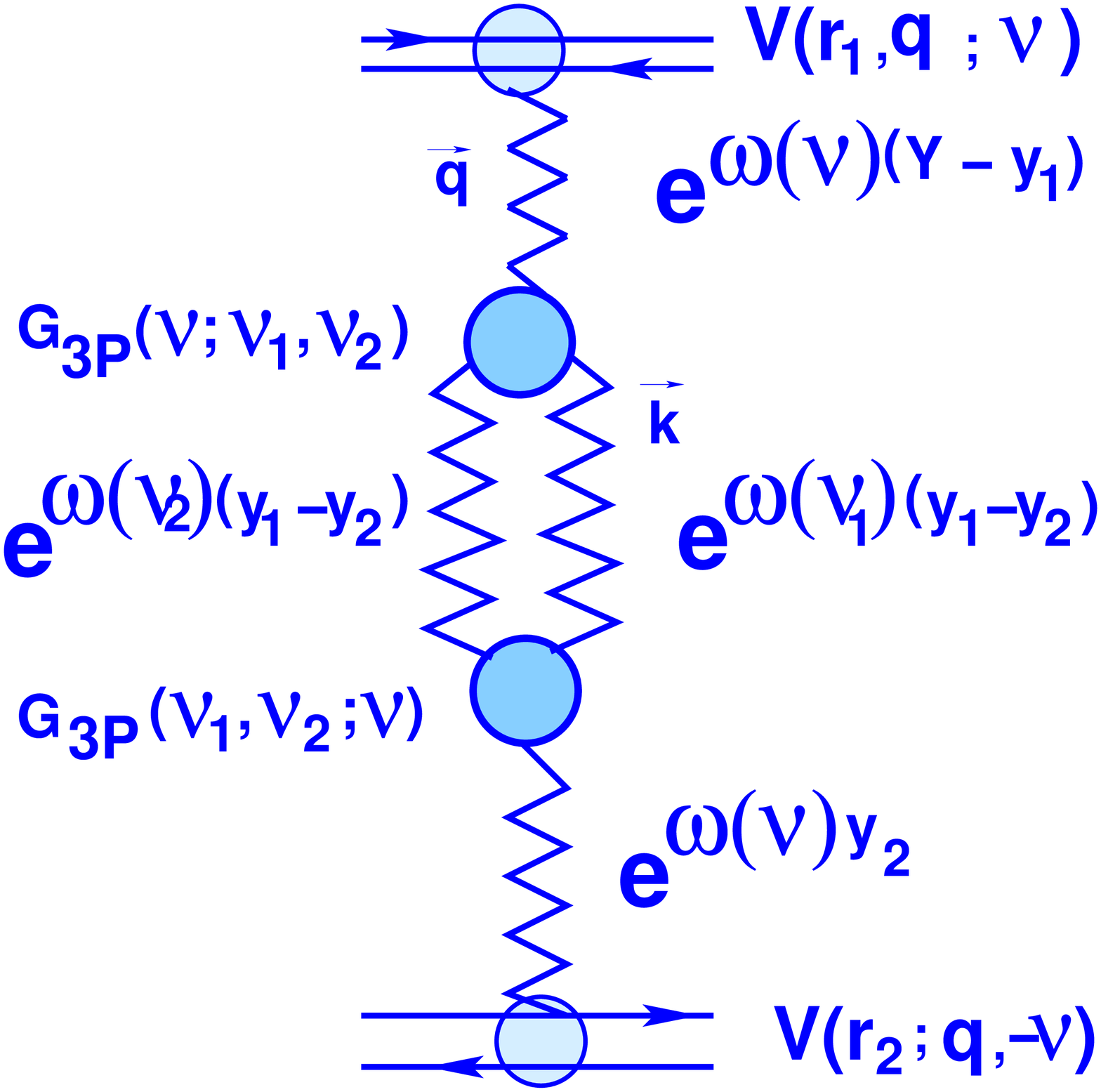,width=90mm}
\end{center}
\end{minipage}
\begin{minipage}{6cm}
\caption{The first enhanced diagram in the Reggeon calculus with the BFKL Pomerons (zigzag lines).}
\label{bfklenh}
\end{minipage}
\end{figure}

In this case,  the typical value for $Y - y_1$ in \eq{BFKLENH} is of the order of 
\beq \label{YEST}
Y\,\,-\,\,y_1\,\,\,\sim\,\,\frac{1}{\as(r_{1,t})}\,\,\gg\,\,1\,\,\,\,\,\,\,\,\,\,
\mbox{while}\,\,\,\,\,\,\,\,\,\,\,\,
y_2\,\,\,\sim\,\,\frac{1}{\as(r_{1,t})}\,\,\,\,\ll\,\,Y \,\,-\,\,y_1
\eeq
If $r_{2,t} \,\,\approx\,\,R$ where $R$ is a typical hadron size, $\as(R) \,\approx\,1$ and $y_2 
\,\,\approx\,\,1$. For such small $y$  we cannot trust the Reggeon like diagrams,  but we can replace 
the 
enhanced diagram of \fig{enhdi} or of \fig{bfklenh} by the `fan' diagram (see \fig{fandi} ). The meaning 
of such a replacement is that in a `fan' diagram the interaction of BFKL  ladders with the target, is a 
subject of the non-perturbative QCD approach. Assuming that the BFKL ladders interacts with the target 
independently (without correlations), we can write the non-linear equation.

Therefore, the non-linear equation is a direct  consequence of two physical ideas: (i) the high energy 
amplitude can be replace by exchange of the BFKL `ladder'; and (ii) in the first approximation the 
correlations can be neglected 
for  the low energy interactions of 
the 
BFKL `ladder' with the target. The last claim has a theoretical 
justification for a  nucleus target \cite{K}, here we do not consider  the interaction with 
nucleus.
\subsection{Generating functional:}
In Ref. \cite{MUUN} A.H. Mueller suggested  separating  the high energy part of `fan' diagrams from the 
low energy part,   by introducing the generating functional. This functional has the form:
\beq \label{GEFU}
Z\left(Y - y,r;[u_i]\right)\,\,\equiv\,\,\sum_{n=1}\,\int
 P_n(Y - y,r; r_1,b_1; r_2,b_2;\, \dots\,r_i,b_i;\,\dots\,r_n,b_n)\,\,
\prod^n_{i = 1} \,u(\vec{r}_i, \vec{b}_i)\,d^2\,r_i\,\,d^2\,b_i
\eeq
where $u(\vec{r}_i)\, \equiv \,u_i$ is an arbitrary function of $r_i$ and 
$b_i$. $P_n$ denotes the probability density of 
 finding  $n$ dipoles with rapidity $y$, with transverse size $r_1,r_2,\, \dots\,r_i\,\dots\,r_n$, and 
with  
impact 
parameters  $b_1,b_2,\, \dots\,b_i\,\dots\,b_n$   with respect to the mother-dipole, 
in the wave function of the fast moving dipole of the size $r$ and rapidity $Y \,>\,y$. The functional 
of \eq{GEFU}  satisfies two conditions:
\begin{enumerate}
\item \quad At $y =Y$,  $P_1 \,=\,\delta^{(2)} (\vec{r} \,-\,\vec{r}_1 )$ with $P_{n \,>\,1} = 0$. This 
condition means that at the beginning of the evolution we have one fast moving dipole or, in other 
words, 
we  sum  the `fan' diagrams which  start with the exchange of one BFKL Pomeron. For the 
functional we have
\beq \label{GEFUIN1}
Z\left( Y - y = 0,r;[u_i] \right)\,\,=\,u(r)\,\,;
\eeq
\item \quad At $u_i = 1$
\beq \label{GEFUIN2}
Z\left( Y - y ,r;[u_i = 1] \right)\,\,=\,1\,\,;
\eeq  
\eq{GEFUIN2} expresses the physical meaning of the functional:the  sum  over all probabilities is equal 
to 
unity.
\end{enumerate}

This functional sums all `fan' diagrams (see \fig{gefunfan} ), and the simple linear functional 
equation 
can be written for it \cite{LL}:

\begin{figure}[htb]
\begin{minipage}{12cm}
\begin{center}
\epsfig{file=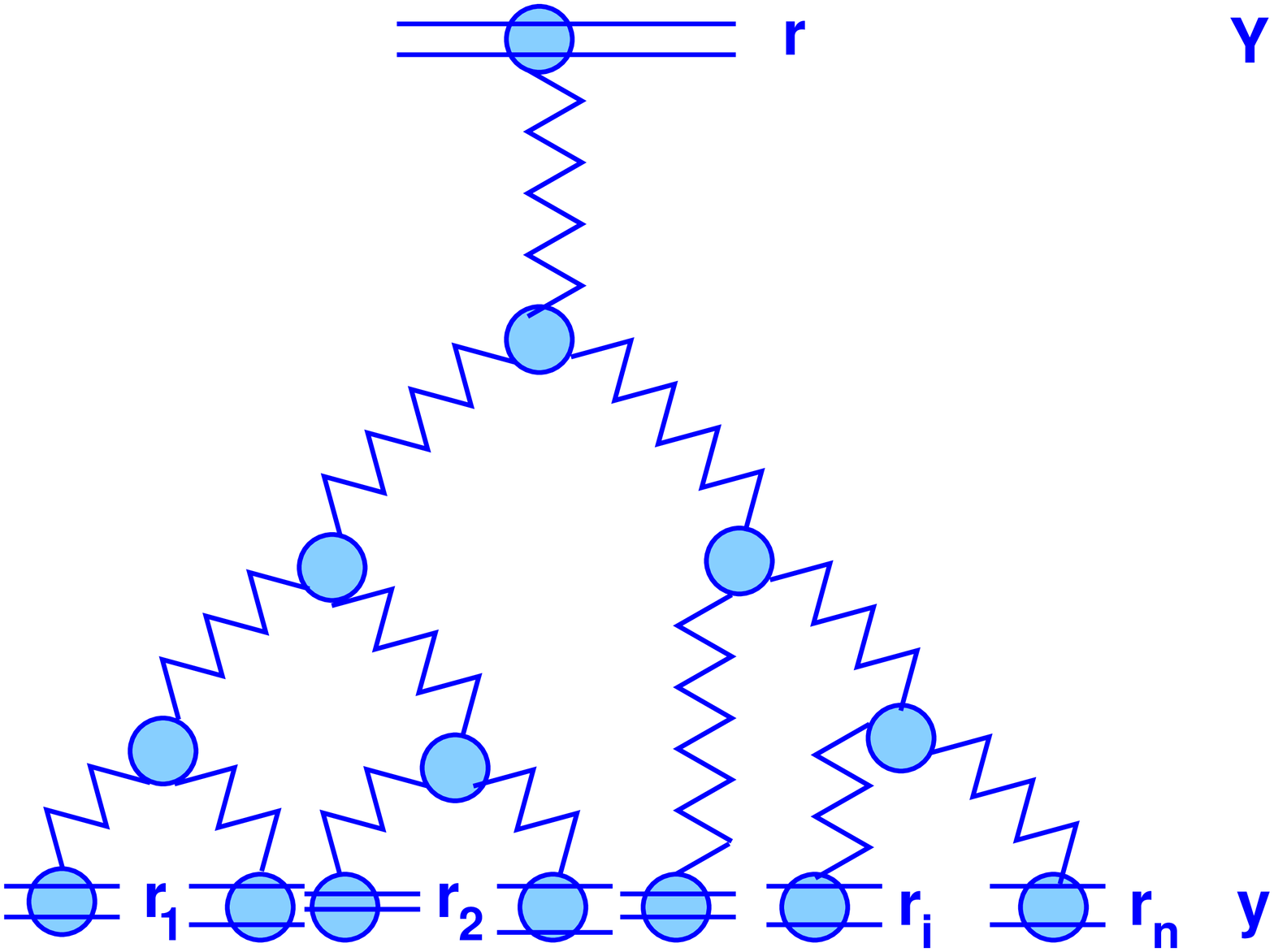,width=110mm}
\end{center}
\end{minipage}
\begin{minipage}{4cm}
\caption{The `fan' diagrams that are summed by the generating functional. Zigzag lines denote the BFKL
Pomerons.}
\label{gefunfan}
\end{minipage}
\end{figure}
\beq \label{LEQGF}
\,\,\frac{\partial Z\left(Y -y,r,;[u_i] \right)}{\partial
\bar{\alpha}_S\,\,
y}\,\,\,=\,\,\,-\,\,\int\,\,d^2 r_i\,u(r_i)\,\omega(r_i)\,\,\frac{\delta}{
\delta
u_i
} \,\, Z\left(Y -y,r;[u_i] \right)\,\,\,
\eeq
$$
+\,\,\int\,\,d^2 \,r_i \,\,d^2\,r'
\,u(r_i)\,u(\vec{r}_i - \vec{r}')\,\, \frac{r'^2}{r^2_i\,(\vec{r}_i - \vec{r}')^2}\,\,
\,\,\frac{\delta}{
\delta
u(r')
} \,\, Z\left(Y -y,r;[u(r'), u_i] \right)
$$
where 
\beq 
\omega(r)\,\,=\,\,\frac{1}{2 \pi}\,\int\,\,d^2\,r'
\,\, \frac{r^2}{r'^2_i\,(\vec{r} - \vec{r}')^2}\,\,,
\eeq
the    notation $\delta/\delta u_i$ is used for the functional derivative.

The physical meaning of each term is clear: the first one describes a probability for the BFKL Pomeron 
to propagate from rapidity $y $ to rapidity $y + d y$ without any decay, while the second term accounts 
 for the 
possibility for decay of one dipole to two dipoles or, in other words,  it takes into account the triple 
BFKL Pomeron vertices. 

A general property of \eq{LEQGF},  is that a solution to this equation can be written as a function of a 
single variable $u(y)$. Rewriting $\partial Z/\partial y $ as $\partial Z/\partial y = \,\int \,d^2 
r\,\left(\delta Z/\delta u(r) \right)\,\left(\partial u(r)/\partial y \right)$ and using the initial 
condition of \eq{GEFUIN1} one obtains a non-linear equation \cite{MUUN}:  
\beq \label{NLEQGF}
\,\frac{d Z\left(Y -y,r;[u_i] \right)}{d
\bar{\alpha}_S\,\,
Y}\,\,=\,\,- \omega(r)\,\,Z\left(Y -y,r;[u_i] \right)\,\,\,
\eeq
$$
+\,\,\int\,\,
d^2\,r'\,\,\frac{r^2}{r'^2\,(\vec{r}\,-\,\vec{r}')^2}\,\,\,Z\left(Y 
-y,r';[u_i]
\right)\,\,Z\left(Y -y,\vec{r}\,-\,\vec{r}');[u_i]
\right)\,\,.
$$

\eq{LEQGF} and \eq{NLEQGF} solve the problem of finding the  probability to produce a number of dipoles 
 of different sizes at rapidity $y$, from the single fast moving dipole at rapidity $Y$. This 
probability is independent of 
 the target. Assuming that all produced dipoles interact with the target independently 
(without correlations),  we can calculate the resulting scattering amplitude for the `fan' diagram (see 
\fig{fandi}). As was proved in Ref.\cite{K}, this amplitude is equal to
\beq \label{NDEF}
N\left(Y,r; [ \gamma(r_i,b_i)] \right)\,\,=\,\,1\,\,\,-\,\,\,Z\left(Y,r,b_t;[\gamma(r_i,b_i) + 
1]\right)\,\,.
\eeq
where $ - \,\gamma(r_i,b_i)$ is the amplitude of the interaction of the dipole of  size $r_i$, at impact 
parameter $b$ and at $y=0$ with the target.

\subsection{The Iancu-Mueller factorization and  enhanced diagrams:}
\begin{figure}[htb] 
\begin{minipage}{12cm} 
\begin{center}
 \epsfig{file=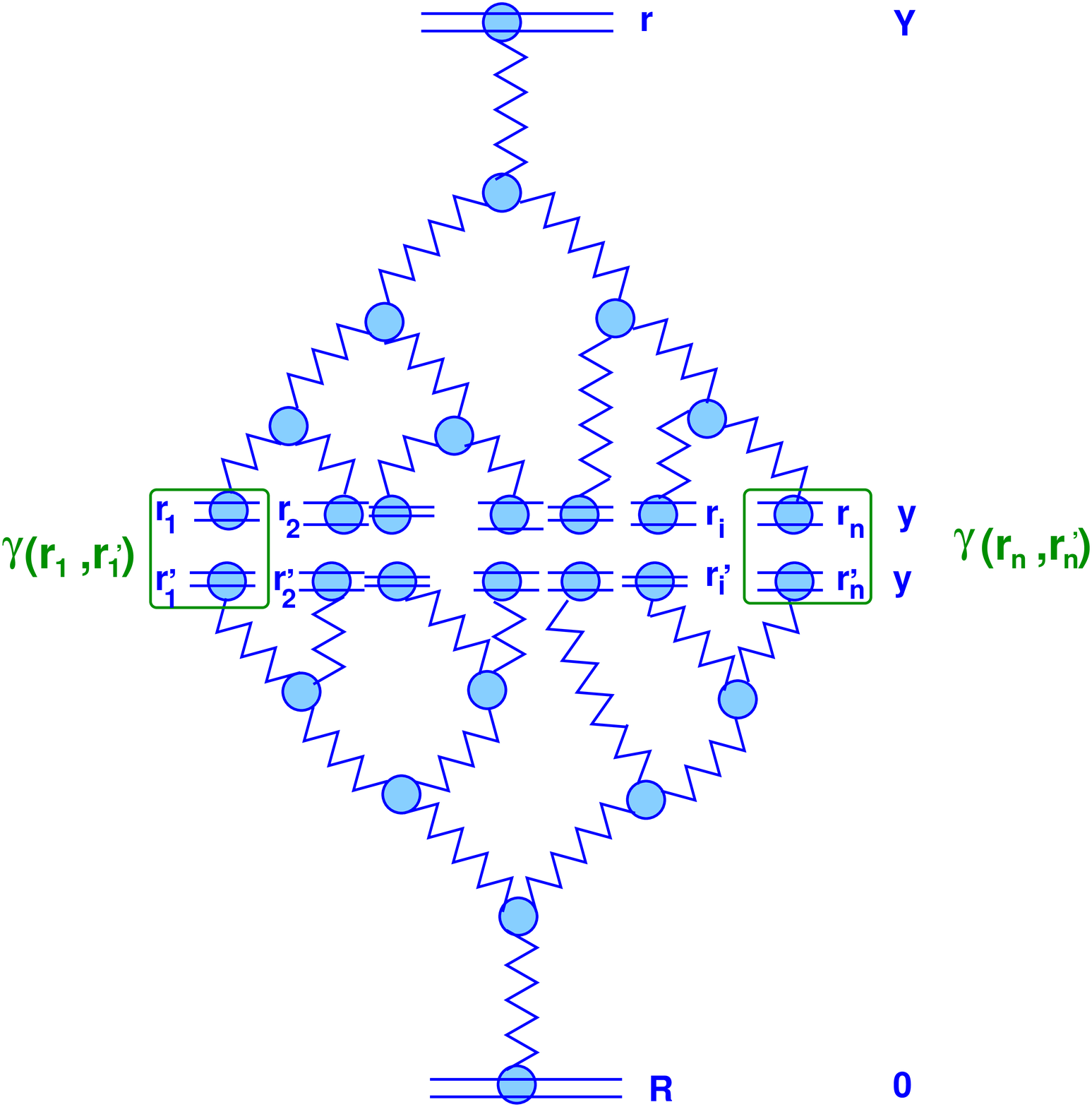,width=110mm}
\end{center} 
\end{minipage}
 \begin{minipage}{4cm} 
\caption{The enhanced diagrams that are summed by the
Iancu-Mueller approach. Zigzag lines denote the BFKL Pomerons.}
 \label{gefunenh} 
\end{minipage}
\end{figure}

The Iancu-Mueller approach to calculations of the enhanced diagrams contribution to the scattering 
amplitude is shown  in \fig{gefunenh}. We need therefore  

 (i) to calculate  the amplitude of production of 
$n$-dipoles of the different sizes, that are produced by both colliding dipoles with sizes $r$ and $R$ 
in 
\fig{gefunenh};

(ii) to multiply the product of these amplitudes by $\gamma(r_i,r'_i)$ for each pair;

(iii) to integrate over all $r_i$, $b_i$  and $r'_i$,  $b'_i$.

 Summing over all possible numbers of the interacting dipoles give us the amplitude.
This procedure leads to the following formula (see \fig{gefunenh} ):
\beq \label{IMFOR}
N\left(r,R,Y;b\right)\,\,\,=\,\,\,\sum^{\infty}_{n=1}\,\,\int\,\,\prod^n_{i=1}\,\, 
d^2b_i\,\,b'_i\,\prod^n_{i=1}\,\, d^2 
r_i\,\,d^2\,r'_i\,\,\,\tilde{\gamma}(r_i,\vec{b} - \vec{b}_i;r'_i,b'_i)\,\,
\eeq
$$
N_n\left(r,Y-y; r_1,\vec{b} - \vec{b}_1; r_2,\vec{b} - \vec{b}_2;\,\dots\,r_i\,\vec{b} - 
\vec{b}_i;\dots\,r_n\,\,\vec{b} - \vec{b}_n \right) 
\,\,N_n\left(R,y; r'_1,b'_1; 
r'_2, b'_2;\,\dots\,r'_i,b'_i\,\dots\,r'_n,b'_n \right)\,\,
$$
where $N_n$ is the general term of expansion of the amplitude given by \eq{NDEF}
$$
N\left(Y,r; [ \gamma(r_i,b_i)] 
\right)\,\,=
$$
\beq \label{NN}
\,\,-\,\,\sum_{n=1}\,\,\prod^n_{i=1}\,\,d^2\,r_i\,d^2\,b_i\,\,\, 
\gamma(r_i,b_i)
N_n\left(r,Y; 
r_1,b_1; 
r_2,b_2\,\dots\,r_i,b_i;\,\dots\,r_n,b_n \right)\,\,.
\eeq
In \eq{NN} we view the scattering  amplitude $N\left(Y,r; [ \gamma(r_i,b_i)] \right)$ as a generating 
functional with respect to arbitrary functions $\gamma(r_i)$. To obtain the 
scattering amplitude for the particular process we have to replace functions  $\gamma(r_i)$ by $  
\tilde{\gamma}(r_i,R)$ where $\tilde{\gamma}(r_i,R) = - A(dipole-target)$ where 
$A$ is the amplitude of the  dipole - target interactions at low energies with sizes $r_i$ and $R$, 
respectively.

Comparing \eq{NDEF} and \eq{NN} one can see that $N_n$ can be calculated as
\beq \label{NNDER}
N_n\left(r,Y,b,r_1,b_1;r_2,b_2;\,\dots\,r_i,b_i;\,\dots\,r_n,b_n 
\right)\,\,\,=\,\,\,\frac{1}{n!}\,\prod^n_{i 
=1}\frac{\delta}{\delta \gamma_i }\,\left(\,\,1\,\,\,-\,\,\,Z\left(Y,r;[\gamma(r_i,b_i) +
1]\right)\,\,\right)|_{\gamma_i = 0}\,\,.
\eeq
As we have discussed, the artificial rapidity $y$ cancels in the product of \eq{IMFOR}.

In principle \eq{IMFOR} solves the problem of calculating the scattering amplitude, if we can find the
expression for $\tilde{\gamma}(r_i,r'_i)$. Comparing \eq{BFKL1} and \eq{BFKLENH} and taking into 
account the
completeness relation of \eq{VV} we obtain that
\beq \label{GAQ}
\tilde{\gamma}(k_i,k'_i;r_i,r'_i)\,\,\,=\,\,(-1)\,\,\delta^{(2)} (\vec{k}_i - 
\vec{k}'_i)\,\,\delta^{(2)} 
(\vec{r}_i -
\vec{r}'_i)\,\,\frac{\bar{\alpha}^2_S\,\pi^3}{N^2_c}\,\frac{1}{D(\nu)}\,\,\frac{1}{r^2_i}
 \eeq
Returning to the impact parameter representation we have \eq{GAQ} in the form
 \beq
\label{GAB}
\tilde{\gamma}(\vec{b}_i, \vec{b}'_i;r_i,r'_i)\,\,\,=\,\,\,(-1)\,\,\delta^{(2)} (\vec{r}_i 
-\vec{r}'_i)\,\,
\delta^{(2)} (\vec{b}'_i  - \vec{b}_i)\,\,\,
\frac{\bar{\alpha}^2_S\,\pi^3}{N^2_c}\,\frac{1}{D(\nu)}\,\,\frac{1}{r^2_i}
\eeq
We recall that $\bas = N_c\,\as/\pi$.

 We can clarify the physical meaning of \eq{GAQ} and \eq{GAB} 
by considering the first term in \eq{NNDER}, which corresponds to the single BFKL Pomeron
exchange. This term is equal
which corresponds to the \beq \label{BFKLSE}
 N\left(Y -y,r,q;[\gamma(r_1)] \right) \,\,\,=\,\,\,\int\,d^2 r_1 N^{BFKL}\left(Y - y,r,r_1;
q\right)\,\,\gamma(r_1) 
\eeq 
Choosing $\gamma(r_1) = \nu$ and $\gamma(r'_1) =\tilde{\gamma}(q;r_1,r'_1)/\nu $,  one can see that
 \eq{NN} could be rewritten in the form 
\bea
 N^{BFKL}\left(Y,r,R; q \right)\,\,&=&\,\,-\,\int \,d^2 r_1 \,d^2r'_1 N\left(Y -
y,r,b;[\gamma(r_1)] \right)\,\,N\left( y,R,q;[\gamma(r'_1)] \right)\, \label{BFKLNN} \\
 &=&\,\,-\int \,d^2 r_1 \,d^2r'_1 N^{BFKL}\left(Y -y,r,r_1; q\right)\,\,\nu\,
N^{BFKL}\left(y,R,r'_1; q \right)\,\frac{\tilde{\gamma}(q;r_1,r'_1)}{\nu} \nonumber \\
 &=&\,\,-
\int \,d^2 r_1 \,d^2r'_1 N^{BFKL}\left(Y -y,r,r_1; q\right)\,\,
N^{BFKL}\left(y,R,r'_1; q\right)\,\tilde{\gamma}(q;r_1,r'_1) \nonumber
\eea
In the last equation we use the completeness of the  BFKL vertex functions $V(r,q,\nu)$ (see \eq{VV} ) 
and 
the explicit form of \eq{GAQ} for $\tilde{\gamma}(q;r_1,r'_1)$, to take the integrals over $r_i$ and
$r'_i$. \eq{BFKLNN} gives  the main formula that allows us to reduce the product 
 of two functional to the exchange of the BFKL Pomeron and, therefore, to the Reggeon -like diagram 
technique.

This formula together with \eq{IMFOR} and \eq{NNDER} leads to the calculation of the scattering amplitude 
in the  Iancu-Mueller approach. However, before performing a calculation using \eq{IMFOR} and \eq{NNDER} 
we 
would like to clarify our approach using  a simple model suggested in Ref. \cite{MUUN}.

\section{The Iancu-Mueller factorization in a toy model}
\begin{figure}[htbp]
\begin{minipage}{9.0cm}
\epsfig{file= 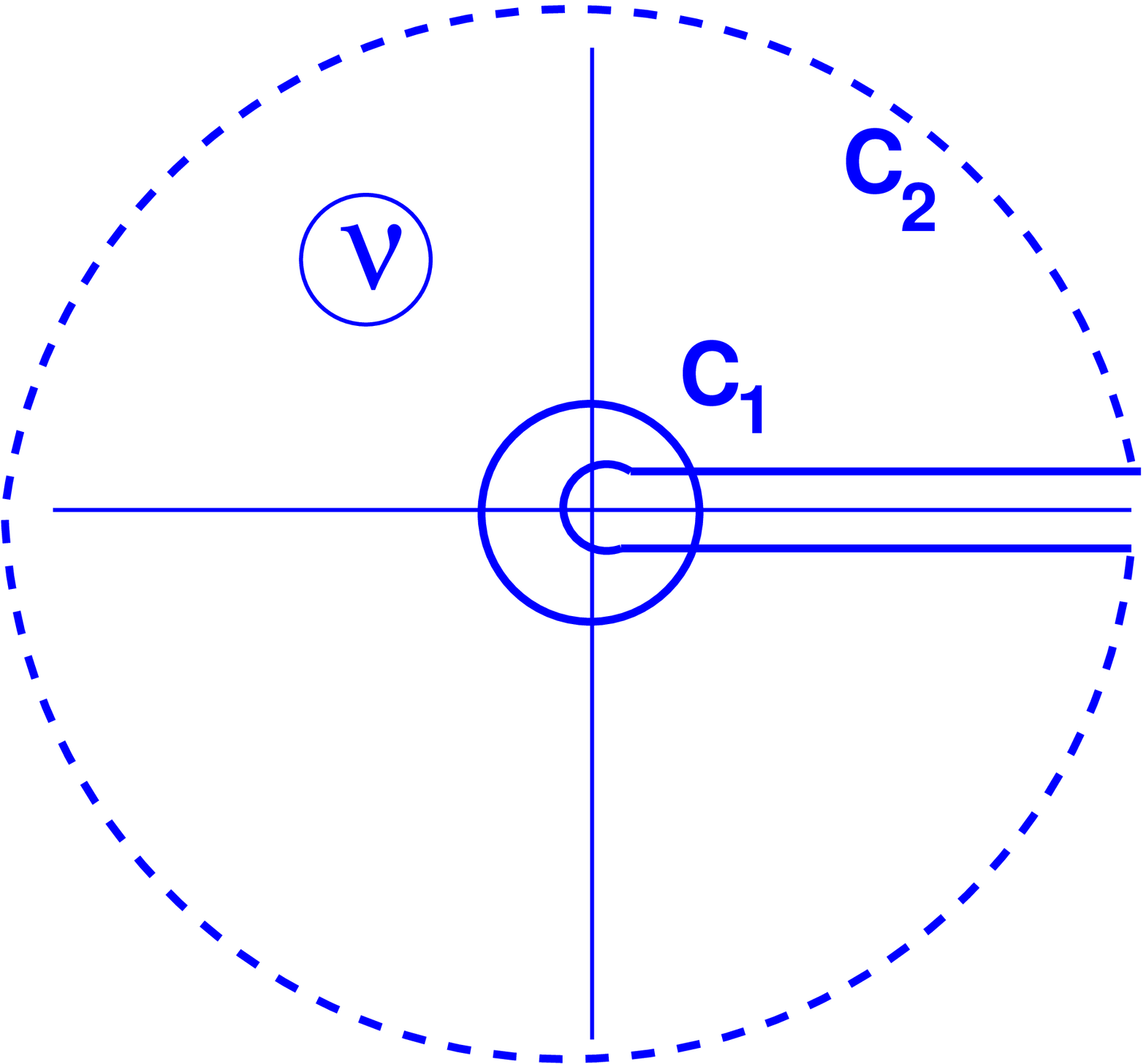,width=80mm}
\end{minipage}
\begin{minipage}{7.0 cm}
\caption{ Complex $\nu$ plane and  contours  for integration over $\nu$ in \eq{IMFORSM} and 
\eq{IMFORFAL} .}
\label{contour}
\end{minipage}
\end{figure}

 We have used  this model in the introduction to illustrate the  main points of our 
approach to the  
Iancu-Mueller factorization. In the toy model, the probability for the dipole to decay in two dipoles is 
a 
constant ($\omega_0$),  which is equal to the probability for a dipole to survive without producing 
another 
dipole. In other words, in the framework of the Reggeon -type diagram technique,  the intercept of 
Pomeron 
is equal to $G_{3P}$ - vertex,  and both are equal to $\omega_0$.  In this model we also neglect the 
fact 
that we have  dipoles of different sizes, and therefore the generating functional of \eq{GEFU} 
reduces  to a generating function
\beq \label{GEF}
Z(Y - y,u) \,\,\,=\,\,\,\sum_{n=1}\,\,\, P_n\,\,u^n
\eeq
with   two initial and boundary conditions:
\bea 
\mbox{At  y = Y :} & Z(Y - y=0,u)\,\,=\,\,u\,\,; \label{GEFIN1}\\
\mbox{At  u =1 :} & Z(Y - y,u=1)\,\,=\,\,1; \label{GEFIN2}
\eea
\eq{LEQGF} reduces to (see Refs. \cite{MUUN,LL})
\beq \label{LEQSMGF}
-\,\,\frac{\partial\,Z(y,u)}{\partial y}\,\,=\,\,-\omega_0
\,\left(u (1 - u)\right)\,\frac{\partial\,Z(y,u)}{\partial u}
\eeq
which has the solution
\beq \label{LEQSMGFSOL}
Z(Y-y,u)\,\,=\,\,\frac{u}{1\,\,+\,\,(\,e^{ \omega_0\,( Y - y)}\,-\,)\,(1 \,-\,u)}\,\,.
\eeq
Using \eq{NDEF} one obtains
\beq \label{NSMSOL}
N(Y-y,\gamma) \,\,=\,\,-\,\,\frac{\gamma\,e^{ \omega_0\,( Y - y)}}{1 \,\,+\,\,\gamma\,\left(\,\,e^{ 
\omega_0\,( Y - y)}\,\,-\,\,1\,\right)}
\eeq

\eq{NSMSOL} together with a natural reduction of \eq{GAB} to
\beq \label{GASM}
\tilde{\gamma} \,\,\equiv\,\,\gamma_{SM}\,\,=\,\,(-1)\,\frac{\bar{\alpha}^2_S\,\pi^3}{N^2_c}
\eeq
allows us to use \eq{IMFOR} to estimate the high energy asymptotic behaviour of the scattering 
amplitude, 
in this simple model. However, we suggest a more compact formula than \eq{IMFOR}, namely,
\beq \label{IMFORSM}
N(Y)\,\,\,=\,\,\,\frac{1}{2\,\pi\,i}\,\oint_{C_1} \,\,\frac{d \nu}{\nu}\,\,N(Y - y, \gamma_{SM} 
\,\nu)\,N(y, 
\frac{1}{\nu})\,\,;
\eeq
where the integration contour is a circle with the unit radius around $\nu = 0$ (see 
contour $C_1$ in \fig{contour}). 
Expanding the functions $ Z$  with respect to 
$\nu$ and $1/\nu$  one can prove in the
case of our simple model  that \eq{IMFORSM} is equivalent to \eq{IMFOR}. 

Taking the integral of \eq{IMFORSM} explicitly we obtain the answer
\beq \label{NSMFIN}
N(Y)\,\,\,=\,\,\,\,\,=\,\,\frac{\frac{\bar{\alpha}^2_S\,\pi^3}{N^2_c}
\,e^{ \omega_0\,Y}}{1 
\,\,+\,\,\frac{\bar{\alpha}^2_S\,\pi^3}{N^2_c}
\,
\left(\,\,e^{\omega_0\,( Y - y)}\,\,-\,\,1\,\right)\,\left(\,\,e^{\omega_0\,y}\,\,-\,\,1\,\right)}
\eeq
where $\gamma_{SM}$ is given by \eq{GASM}.  On can see that \eq{NSMFIN} leads to
\beq \label{NSMAS}
\lim_{Y\, \gg\,1} 
N(Y)\,\,\rightarrow\,\,1\,\,-\,\,\frac{\bar{\alpha}^2_S\,\pi^3}{N^2_c}
\,e^{-\omega_0\,Y}\,\,+\,\,O 
\left(\,\frac{\bar{\alpha}^2_S\,\pi^3}{N^2_c}\,e^{-\omega_0\,Y/2}\,\right)
\eeq

\section{High energy asymptotic behaviour of the scattering amplitude}
In this section we return to the general expression for the  Iancu-Mueller factorization given by 
\eq{IMFOR} 
and \eq{NNDER}. We would like to calculate the high energy asymptotic behaviour based on our  
experience 
with the simple model, that has been discussed in the previous section. 

We start with discussion of dipole-dipole amplitude in the simplified model for the BFKL kernel, 
namely, assuming that 
\beq \label{SMK}
\omega(\gamma)\,=\,\frac{\as \,N_c}{\pi}\,\left\{\begin{array}{c} 
\,\,\,\,\,\,\frac{1}{\gamma}\,\,\hspace*{1cm}\mbox{for}\,\,r^2\,Q^2_s\,<\,1\,; \\ \\ \\
\,\frac{1}{1\,-\,\gamma}\,\,\hspace*{1cm} \mbox{for}\,\,r^2\,Q^2_s\,>\,1\,; 
\end{array} \right.
\eeq
instead of the general BFKL kernel given by \eq{OMEGA}. In \eq{SMK} $Q_s$ denotes the saturation 
momentum.  It is shown in Ref.\cite{LT} that \eq{SMK} sums 
double log contributions  of the order of $ \left(\as Y \ln(r^2\,\Lambda^2) \right)^n$ in the kinematic 
region of 
perturbative QCD, namely,  $r^2\,Q^2_s\,<\,1$, while in the saturation region ($r^2\,Q^2_s\,>\,1$)  it 
takes into account  large terms  of $ \left(\as  \ln(r^2\,Q^2_s) \right)^n $ -type. In section 4.3 we 
will return to discussion of the general kernel of \eq{OMEGA}.

\begin{boldmath}
\subsection{Fixed $\as$:}
\end{boldmath} 
\subsubsection{ Generating functional for the Balitsky - Kovchegov scattering amplitude:} 
We  calculate the high energy asymptotic behaviour of the scattering amplitude using our 
experience with the simple toy model, and the fact that amplitude $N\left(Y,r,b; [\gamma(r_i)] \right)$ 
has been found in Ref. \cite{LT}. We will show that a direct generalization of \eq{IMFORSM} leads to 
the high energy  amplitude that has been proposed in the  Iancu-Mueller paper \cite{IM}, namely,
\beq \label{IMAS}
N(Y,r_1,r_2;b)\,\,\,\Longrightarrow\,\,\,1\,\,-\,\,e^{- 
\frac{1}{2}\,c\,(Y\,-\,Y_0)^2}
\eeq
where coefficient $c$ determines the asymptotic behaviour of the solution to the Balitsky-Kovchegov 
non-linear equation. This behaviour was found in  Ref. \cite{LT} and it has the form
\beq \label{LTAS}
N^{BK}(Y,r_1;b)\,\,\,\Longrightarrow \,\,\,\,1 \,\,-\,\,e^{- 
\,c\,(Y\,-\,Y_0)^2}
\eeq
with $c =2\, \bas^2$. 
 
The first problem that we need to solve,  is to build the generating functional for the amplitude (see 
\eq{NN}), based on the solution given in Ref. \cite{LT}.

~

\begin{boldmath}
\centerline{$r_t \,\,\,<\,\,\,1/Q_s(Y - y, b)$}
\end{boldmath}

~

 As was shown in this paper, the solution for the 
short distances ($r_t\,Q_s\,\,<\,\,1$) is the single BFKL Pomeron, and, therefore, the generating 
function for the scattering amplitude at  these distances is
\beq \label{GFNSD}
N_{sd}\left(Y - y ,r_1,b;[\gamma(r_i)] \right)\,\,\,=\,\,\,\int\,\,d^2 r_i \,\,\gamma(r_i)\,\,  
N^{BFKL}((Y - 
y ,r_1,r_i,b)
\eeq

~

\begin{boldmath}
\centerline{$r_t \,\,\,>\,\,\,1/Q_s(Y - y, b)$}
\end{boldmath}

~

For long distances, the solution for the interaction of the dipole ($r_1$) with the dipole ($r_i$)  can 
be written in the form \cite{LT}:
\beq \label{SOL1}
N\left(Y - y ,r_1,r_i; b \right)\,\,=\,\,1\,\,\,-\,\,\,e^{- \phi(z)}
\eeq
where
\beq \label{ZR}
z\,\,=\,\,\ln\left(r^2_1\, Q^2_s \right)\,\,=\,\,4 \bas \,(Y - y) 
\,\,-\,\,\ln(r^2_i/r^2_1) 
\,\,+ \,\,2 \,\beta(b)
\eeq
for the kernel of \eq{SMK}.
 $\beta(b)\, = \,2 \,\ln S(b)$ in this approach, where
$S(b)$ is given by the impact parameter dependence of  the Born approximation amplitude for this 
interaction (see Ref. \cite{LT} for details).

Function $\phi$ is defined \cite{LT}  by
\beq \label{PHI}
z\,\,\,=\,\,\,\sqrt{2}\,\int^{\phi}_{\phi_0}\,\,\frac{ d\,\phi'}{\sqrt{ \phi'\,\, + \,\, 
e^{-\phi'}\,\,-\,\,1}}
\eeq
where the value of $\phi_0$ should be found by  matching the solution with \eq{SOL1} at $r_1 = 1/Q_s$.

Assuming that $\phi_0$ is small we obtain from \eq{PHI} that for $z < 1$
\beq \label{MATCH1}
N\left(Y - y ,r_1,r_i; b \right)\,\,=\,\,\phi_0 \,e^{\frac{1}{2}\,z}
\eeq
Comparing  \eq{MATCH1} with the functional of \eq{GFNSD} and recalling that $N^{BFKL} \,\,\rightarrow\,
exp[ \frac{1}{2} z]$ for $r_1$ approaching $1/Q_s$, we obtain that in the region of long distances, but
for $r_1$ close to $1/Q_s$, that 
the 
generating functional for 
scattering amplitude has the   form:
\beq \label{MATCH2}
N_{ld}\left(Y - y ,r_1,b;[\gamma(r_i)] \right)\,\,\,=\,\,\,\int\,\,d^2 r_i 
\,\,\gamma(r_i)\,\,e^{-\frac{1}{2}\,\ln(r^2_i)}
\eeq

Using \eq{MATCH2} and \eq{PHI} we can see  that   in the region of large $z$ ($Y - y 
\,\,\gg\,\,\,\ln(1/r^2)\,\,\gg\,\,1$) the generating functional is
\beq \label{GFNLD}
N_{ld}\left(Y - y ,r_1,b;[\gamma(r_i)] \right)\,\,\,=\,\,\,1\,\,-\,\,e^{- 
\frac{1}{2}\,\left(\,\,\frac{z}{2} + \Phi[\gamma(r_i)]\,\,\right)^2}
\eeq
with the functional $\Phi([\gamma(r_i)])$ defined as 
\beq \label{PHIGA}
\Phi([\gamma(r_i)]) \,\,\,=\,\,\, \ln \left( \int \,\,d^2 \,r_i \,e^{ - \frac{1}{2} 
\ln(1/r^2_i)}\,\,\gamma(r_i) \right)
\eeq
where $z$ is defined by \eq{ZR}.

The accuracy of \eq{GFNLD} is not very high and we cannot guarantee the value 
of the constant in front of the exponent in \eq{GFNLD}.

\subsubsection{Dipole-dipole scattering amplitude in Iancu - Mueller approach:}

To calculate the dipole-dipole scattering amplitude we would like to generalize \eq{IMFORSM}. We replace 
$\gamma(r_i) $ by $ \nu$, while we choose $\gamma(r'_i) = \tilde{\gamma}(r_i,r'_i)/\nu$ (see \eq{GAB} ).

Using these functions $\gamma$ we can rewrite \eq{IMFOR} in the form of \eq{IMFORSM}, namely,
\beq \label{IMFORFAL}
N\left(r,R,Y;b\right)\,\,\,=\,\,\,\frac{1}{2\, \pi\, i}\,\oint_{C_2}\,\frac{d \nu}{\nu}\,
N\left(Y - y ,r,b;[\nu] \right)\,\,N\left(y ,R,b;[\tilde{\gamma}(r_i,r'_i)/\nu] \right)
\eeq
where we deform contour $C_1$ to contour $C_2$ (see \fig{contour} ).
In \eq{IMFORFAL} we use $r$ and $R$ instead of $r_1$ and $r_2$ denoting the sizes of colliding dipoles.

Substituting \eq{GFNLD} in this equation we obtain the resulting expression 
\beq \label{IMFORR}
N\left(r,R,Y;b\right)\,\,\,=
\eeq
$$
\,\,\,\frac{1}{2\, \pi\, i}\,\oint_{C_2}\,\frac{d \nu}{\nu}\,\,\left(\,\, 1
\,\,\, +\,\,\,e^{- \frac{1}{2}\,\left(\,\,\left(\,\,\frac{z\,(Y-y,r)}{2} + \Phi[\nu]\,\,\right)^2
\,+\,\left(\,\,\frac{z\,(y,R)}{2} + 
\Phi[\tilde{\gamma}(r_i,r'_i)/\nu]\,\,\right)^2\right)}\,\,\right)
$$ 
In
\eq{IMFORR} we denote the sizes of interacting dipoles by $r$ and $R$ instead of $r_1$ and $r_2$. As 
we have
seen in toy model example, the
contour of integration is situated between all singularities of two amplitudes in \eq{IMFORR}.
 Due to this fact the contributions such as 
$$- e^{-
\frac{1}{2}\,\left(\,\,\frac{z\,(Y - y, r)}{2} +
\Phi[\gamma(r_i)]\,\,\right)^2}\,\,\,\,\,\,\,\,\,\,
 \mbox{and}
\,\,\,\,\,\,\,\,\,\,
- e^{-
\frac{1}{2}\,\left(\,\,\frac{z\,( y, R)}{2} +
\Phi[\gamma(r_i)]\,\,\right)^2}
$$
are equal to zero since they have singularity outside of contour $C_2$. Since the integral over the 
large 
circle (see dotted line in \fig{contour}) is equal to zero, \eq{IMFORR} can be rewritten as follows
\beq \label{IMFORI}
N\left(r,R,Y;b\right)\,\,\,=
\eeq
$$
\,\,\,1\,\,+\,\,\frac{1}{\pi}\,\left(1 - e^{2 \pi^2} \right)\,\,
\int^{+ \infty}_{- \infty}\,\,d\,l\,\,e^{- \frac{1}{2}\,\left(\,\,\left(\,\,\frac{z\,(Y-y,r)}{2} 
+ \,\,l\,\,+\,\,\chi\,\,\right)^2
\,+\,\left(\,\,\frac{z\,(y,R)}{2} \,\,-\,\,l\,\,+\,\, \chi[\tilde{\gamma}(r_i,r'_i)]\,\,\right)^2 
\,\,\right)}
$$
We are not sure of the value of the coefficient in front of the integral, 
but we wrote  it explicitly,  to show that the second term is  negative. In \eq{IMFORI} we 
extract 
the variable $\nu$ from both $\Phi$'s and use the notation $l \,\equiv\,\ln \nu$. Taking  the 
integral over $l$ explicitly one obtains
\beq \label{IMFORF}
N\left(r,R,Y;b\right)\,\,\,=\,\,\,1\,\,\,-\,\,e^{-\,\frac{1}{16}\,z^2(Y,r,R)}
\eeq
where
\beq \label{ZRR}
z(Y,r,R)\,\,\,=\,\,\,4 \,\bas\,Y\,\,\,-\,\,\,\ln(R^2/r^2)
\eeq

 \eq{IMFORF} reproduces the result of the Iancu and Mueller paper \cite{IM} ( with numerical 
coefficients of our simplified model for the BFKL kernel given by \eq{SMK})  
and it 
does not contain any variable related to the  fictional dipoles with 
rapidity $y$ as well as rapidity $y$ itself. 

The sum of `fan' diagrams leads to the answer \cite{LT}
\beq \label{FANF}
N_{fan}\left(r,R,Y;b\right)\,\,\,=\,\,\,1\,\,\,-\,\,e^{-\,\frac{1}{8}\,z^2(Y,r,R)}
\eeq
which is much closer to unity than the correct answer given by \eq{IMFORF}. In the case of
 fixed QCD coupling our calculations as well as the Iancu and  Mueller ones, show that 
the 
non-linear BK equation is not able to describe the physics of the  dipole-dipole interactions, even when 
dipoles have quite  different sizes (say $R\,\,\gg\,\,r$).

 \begin{boldmath}
\subsection{Running \,$\as$:}
\end{boldmath}
It is not theoretically clear how to include the running QCD coupling in this approach, mostly because 
we do not know how to write the BFKL kernel for running QCD  coupling. However, for short distances in 
the pQCD phase ($r^2\,Q^2_s\,\leq\,1$) we know that $\as$ in the BFKL equation depends on the size of 
the scattered dipole \cite{GLR,MUT,MU02}. This dependence leads to the saturation scale $Q_s(Y) 
\,\,\propto\,\,e^{\sqrt{c Y}}$ and to the geometric scaling behaviour versus  the variable $r^2Q^2_s$ 
(at 
least within the semi-classical accuracy)  \cite{GLR,MUT}.

 Inside 
the  saturation region  ($r^2\,Q^2_s\,\gg\,1$) we can assume that the running $\as$ is frozen at the 
saturation scale 
\cite{BKL}.
This assumption appears  natural from the point of view that physics of saturation is determined by one 
scale: the saturation momentum \cite{GLR,MUQI,MV}. Indeed, it was shown in Ref. \cite{BKL},  that this 
assumption leads to the geometrical scaling \cite{GSC} in the saturation region, making our hypothesis 
 self-consistent and providing the natural matching with the pQCD region.

In the framework of this assumption we obtain the same solution  given by \eq{SOL1} but with a new 
 variable $z$, namely
\beq \label{ZRUN}
z \,\,=\,\,\ln \left(Q^2_s \,\,r^2_1\right)\,\,=\,\,2 \bas(Q^2_s)\,(Y 
\,-\,y)\,\,-\,\,\ln(r^2_i/r^2_1)\,\,+\,\,2 \beta(b)\,\,.
\eeq
In \eq{ZRUN} we use the expression for the saturation scale in the case of running $\as$  obtained in 
Refs. \cite{GLR,MUT,BKL}. We would like to recall that in Refs. \cite{GLR,MUT,BKL} instead of factor 2
in the second line stands $(1/2)\,d \chi(\gamma)/d \gamma |_{\gamma = \gamma_{cr}}$, where $\gamma_{cr}$ is 
the solution to the equation $\chi(\gamma_{cr})/(1 - \gamma_{cr}) = - d \chi(\gamma)/d \gamma |_{\gamma = 
\gamma_{cr}}$. However, in DLA the BFKL kernel $\chi(\gamma)$ is so simple that $\gamma_{cr} = 1/2$ and  
$(1/2)\,d \chi(\gamma)/d \gamma |_{\gamma = \gamma_{cr}} = 2$. In section 4.3 we will discuss the general 
form of the BFKL kernel and all these numerical factors will reappear in our calculations.

Finally, after performing the same calculation as has been discussed above, we obtain the same formula 
of \eq{IMFORF} but with
\bea 
Z(Y,r,R)\,\,&=&\,\,2 \bas(Q_s(Y - y))\,(Y - y) \,\,+\,\,2 \bas(Q_s(y))\, y \,\,-\,\,\ln(R^2/r^2) 
\nonumber \\
  &=&\,\,\frac{8 \,N_c}{b}\,\left(\,\sqrt{Y - y}\,\,\,+\,\,\,\sqrt{y}\,\right)  
\,\,-\,\,\ln(R^2/r^2)\,\,; 
\label{ZRUNR} 
\eea

The minimum of $Z(Y,r,R)$ occurs  at $y=0$ or  at $y = Y$. The case $y=0$ in our notation corresponds to 
the `fan' diagrams of \fig{fandi}-a type (see also \fig{gefunfan}), if we assume that $R > r$. We indeed 
assumed this, considering $\ln(r_i/r)\,> 0$ and $\ln(R/r'_i) > 0$  when we tried to justify the
generating functional approach  for the `fan' diagrams. As has been discussed in section 2.2 (see also 
\cite{GLR} 
), the selection of the `fan' diagrams was  defined for the dipole sizes less than $1/Q_s$. 

Therefore, only the solution for $y =0$ can match the amplitude at $ r < 1/Q_s$.
Finally, our calculations support the principal idea of the non-linear equation that in the case of 
running QCD coupling  it describes the 
interaction of the dipoles with  different sizes. However, this conclusion is based on an additional 
assumption that the running QCD coupling is frozen on the saturation scale.

As was argued in Ref. \cite{MU02},  in the approach given by \eq{SMK}, it seems  
 natural 
to assume that 
$\as$ depends on the size of produced dipole. Indeed, in this case we expect that the size of 
the produced dipole ($r'$ or $|\vec{r} - \vec{r}'|$ in \eq{NLEQGF} ) is smaller than the size of 
the scattered dipole but larger than $1/Q_s$ inside the saturation domain \cite{LT,MU02}. Therefore, 
we can replace 
\beq \label{RUNKER}
\as \int \,\, \frac{r^2}{r'^2 \,(\vec{r} - 
\vec{r}')^2}\,\,\rightarrow\,\,\pi\,\int^{r^2}_{1/Q^2_s}\,\as(r')\,\frac{d r'^2}{r'^2}\,\,\,+\,\,
\pi\,\int^{r^2}_{1/Q^2_s}\,\,\as(|\vec{r} - \vec{r}'|)\,\,\frac{d\,(\vec{r} - 
\vec{r}')^2}{(\vec{r} - \vec{r}')^2}\,\,.
\eeq

Introducing a new function $\tilde{N}(\xi,y)\,\,=\,\,\int^{\xi_s}_{\xi}\,\,\as(\xi') \,d \xi' 
\,\,N(xi',y)$ where $N$ is the dipole scattering amplitude and $\xi \,\,=\,\,\ln(1/(r^2\Lambda^2))$ 
and $\xi_s \,\,=\,\,\ln(Q^2_s/\Lambda^2)$, we 
can rewrite the Balitsky-Kovchegov equation in the following way
\beq \label{RUNEQ}
-\,\,\frac{d^2 \tilde{N}(\xi,y)}{d y \,d\,\xi}\,\,=\,\,\frac{2\,C_F}{\pi}\,\left( 
\,\as(\xi)\,\tilde{N}\,\,+\,\,\tilde{N}(\xi,y)\,\,\frac{d \tilde{N}(\xi, y)}{ d \xi} \right)
\eeq
which is a direct generalization of  Eq.(2.13) in  Ref. \cite{LT}. Replacing $N(\xi,y) $ by function 
$\phi(\xi, y)$ using $N = 1 - e^{- \phi(\xi,y)}$ one can see that \eq{RUNEQ} reduces to
\beq \label{RUNEQ1}
\frac{d \,\phi}{d y} \,\,=\,\,\frac{2 \,\as(\xi)\,C_F}{\pi}\,\tilde{N}
\eeq
Assuming that $\phi$ increases moving inside of  the saturation region , one can see that $\tilde{N}$ 
approaching 
\beq \label{NT}
\tilde{N}\,\,\,\longrightarrow\,\,\,\frac{4 \pi}{b} \,\ln\left( \frac{\xi_s}{\xi} \right)
\eeq
Substituting \eq{NT} in \eq{RUNEQ1} and integrating with respect to $y$ using the explicit expression 
for $\xi_s$ ($ \xi_s\,\, =\,\, \sqrt{\frac{16\,N_c}{\pi \,b}\,\,y}$) we obtain
\beq \label{RUNSOL}
\phi(\xi,y) \,\,=\,\,\frac{8 \,\as(\xi)\,C_F}{b}\,\,\{\,y\,\, \left(\frac{1}{2}\,(\ln y - 1) + \ln 
\sqrt{\frac{16\,N_c \,}{\pi \,b}} \,-\,\,\ln \xi \right)\,\,+\,\,C(\xi)\,\}
\eeq
Function $C(\xi)$ in \eq{RUNSOL} can be found from the condition that $\phi(\xi = \xi_s) = Const$ . 
The final answer has the form \footnote{This solution is very similar to the solution obtained in 
Ref.\cite{MU02} but it has an extra factor $\as(\xi)$. It should be stressed that only with this 
factor this solution could be matched with the geometrical scaling solution in the pQCD region 
($r^2Q^2_s <1$ ).}
 \beq \label{RUNSOLF}
\phi(\xi,y)\,\,=\,\,\frac{\as(\xi)}{2\, \pi}\,\left( \xi^2_s\,(\ln(\xi_s/\xi) - 1) + \xi^2 \right)
\eeq
One can see that this solution does not show the geometric scaling behaviour. However, if $\xi$ is 
close to $\xi_s$ \eq{RUNSOLF}  degenerates into  geometrical scaling behaviour within accuracy 
$\ln(Q^2_s\,r^2)/\ln(Q^2_s/\Lambda^2)$. Therefore, this solution  is not worse that the one 
discussed before 
with the  exact geometrical scaling behaviour. 

Repeating the same calculation as in section 4.1 but with the solution given by \eq{RUNSOLF} we obtain 
the behaviour 
\beq \label{RUNMI}
N\left(r,R,Y;b \right)\,\,=\,\,1\,\,-\,\,e^{ - \phi(Y-y,r) - \phi(y,R)}
\eeq
One can see that minimum of the sum of $\phi$'s occurs at 
\beq \label{MIN}
\xi^2_s(y_{min}) = 
\xi^2(r)\,\left(\frac{\xi^2_s(Y)}{\xi^2(R)}\right)^{ \frac{\as(r)}{\as(R)}}
\eeq
where we used  obvious notation $\xi_s(Y) \,\,=\,\,\ln(Q^2_s(Y)/\Lambda), \xi(r) 
\,=\,\ln(1/(r^2\Lambda^2))$ and so on. Assuming that $\as(r) \,\ll\,\,\as(R)$ we obtain
the following behaviour for the scattering amplitude in Iancu - Mueller approach:
\beq \label{RUNMIF}
 \phi(Y-y,r) + \phi(y,R)\,\,\longrightarrow\,\, \phi(Y,r) \,+\, \frac{\as(\xi)}{2\, \pi}\,\left( -\,
\xi^2_s(y_{min})\, + \xi^2(R) \right)\,\,,
\eeq
where $y_{min}$ is defined by \eq{MIN}.

One can see that $y_{min}$ depends on $Y$ and, therefore, the asymptotic behaviour of the scattering 
amplitude cannot be calculated just using the Balitsky-Kovchegov equation on the contrary to our example 
with frozen $\as$.
On the other hand, for 
$\as(r)\,\ll\,\as(R)$ $y_{min}$ is almost constant and the influence of the target is rather small.

These two examples illustrate the importance of understanding  the argument in QCD the running  coupling 
constant 
in the framework of the non-linear equation. In our simple model with \eq{SMK} for the BFKL kernel the 
second example looks more reliable.

\subsection{High energy scattering amplitude with the general BFKL 
kernel}

The estimates discussed above are  based on the solution to the non-linear 
Balitsky-Kovchegov equation given in Ref. \cite{LT}. In this section we are going to show that this 
solution cannot give a correct high energy asymptotic behaviour. In this section we use  the  full BFKL 
kernel and solve the Balitsky-Kovchegov equation  in the saturation region.

\subsubsection{The solution to the non-linear equation in the saturation 
region:}
 To  search for  the solution in the saturation region, we use  several ideas and 
technical methods  that have 
been discussed in Refs. 
\cite{LT,BKL}.

1.  It is  useful to consider the non-linear equation in a mixed representation, fixing 
impact parameter $b$,  but introducing the transverse momenta as conjugated variables to the  dipole 
sizes. 
The relations  between these two representations are given by the following equations \cite{BKL,MP}
\bea 
N(r,y;b)\,\,\,&=&\,\,\,\,r^2\,\int^{\infty}_0
 \,k d k \,J_0(k\,r)\,\,\tilde{N}(k,y;b)\,\,;\label{MR1}\\
\tilde{N}(k,y;b)\,\,\,&=&\,\,\,\,\int^{\infty}_0\,\frac{d 
r}{r}\,\,J_0(k\,r)\,\,N(r,y;b)\,\,; \label{MR2}
\eea

2.  In this representation the non-linear equation reduces to the form:
\beq \label{NLEQMR}
\frac{\partial\,\tilde{N}(k,y;b)}{\partial 
y}\,\,\,\,=\,\,\,\,\bas\,\left(\,\chi(\hat{\gamma}(\xi))\,\tilde{N}(k,y;b)\,\,\,-
\,\,\,\tilde{N}^2(k,y;b)\,\right)
\eeq
where $\chi(\hat{\gamma}(\xi)$ is an operator defined as
\beq \label{GAMMA}
\hat{\gamma}(\xi)\,\,\,=\,\,1\,\,\,+\,\,\,\frac{\partial}{\partial\,\xi}
\eeq
where $\xi \,\,=\,\,\,\ln(k^2\,k'^2\,b^4)$,  and   $k$ and $k'$  are the  conjugated variables to the 
dipole 
sizes 
of the projectile and the  target, $b$ is an impact parameter which we assumed to be large in 
\eq{NLEQMR}.

3.   We expect there to be   geometrical scaling behaviour of the scattering amplitude in the 
saturation domain. 
It means that $\tilde{N}(k,y;b)$ is a function of the single variable 
\beq \label{ZMR}
z\,\,\,=\,\,\ln(Q^2_s(y,b)/\Lambda^2)\,\,\,-\,\,\,\,\xi(b)\,\,=\,\,\,\bas \,\frac{\chi(\gamma_{cr})}{1 
\,-\,\gamma_{cr}}\,\,(\,y\,\,-\,\,y_0\,)\,\,\,-\,\,\,\,\xi(b)\,\,;
\eeq  
where $\gamma_{cr}\,\,\approx\,\,0.37$ is a solution of  the equation \cite{GLR,MUT}
\beq \label{GAMMACR}
\frac{d\,\chi ( \gamma_{cr})}{d\,\gamma_{cr}}\,\,\,=\,\,\,- \,\frac{\chi ( 
\gamma_{cr})}{1\,\,-\,\,\gamma_{cr}}\,\,;
\eeq

4. Introducing a function $\phi(z)$  we are looking for the solution of the equation in the form 
\beq \label{PHIMR}
\tilde{N}(z)\,\,\,=\,\,\,\h \,\int^{z}\,\,d\,z'\,\,\left(\,1\,\,\,-\,\,\,e^{ - 
\,\phi(z')}\,\right)\,\,;
\eeq

5.  We assume that function $\phi$ is a smooth function, such that $\phi_{z z} 
\,\,\ll\,\,\phi_z\,\phi_z$ where we denote $\phi_z = d \phi/d z$ and $\phi_{z z} = d^2 \phi/(d z)^2$.
This property allows us to rewrite 
\beq \label{SCDER}
\frac{d^n}{(d z)^n}\,e^{- \phi(z)}\,\,\,=\,\,\,( - \phi_z)^n \,e^{- \phi(z)}
\eeq
Substituting in  \eq{NLEQMR}  $\tilde{N}$ in the form of \eq{PHIMR} and replacing $y$ by $\hat{z}$ 
we obtain 
\beq \label{NLEZFS}
\bas \,\frac{\chi(\gamma_{cr})}{1
\,-\,\gamma_{cr}}\,\,\frac{d^2 \tilde{N}(z)}{(d z)^2}\,\,=\,\,
 \bas 
\left(
\,\,[ f\,\chi(1 -f)\,\,-\,\,1]\,\tilde{N}(z)\,\,+\,\,\tilde{N}(z)\,e^{ -\phi}\,\,\right)
\eeq
after taking the  derivative with respect to  $z$ on  both sides of \eq{NLEQMR}. 
$ f$ denotes $f = d/d z$ in \eq{NLEZFS}. An important property of function 
$f\,\chi(1 -f)\,\,-\,\,1$,  is the fact that at 
small $f$ it has the expansion that starts \footnote{It should be stressed that the 
simplified 
model for $\chi(1 -f) = \frac{1}{f (1 - f)}$ does not have this property. This is an explanation why in 
Ref. \cite{LT} where this model was used, the solution was missed  as  we will discuss below.}  from 
$f^2$.  Using 
\eq{SCDER} one can see that the first term on 
r.h.s. of \eq{NLEZFS} is proportional to $e^{ -\phi}$.  Canceling $\,\bas\,e^{ -\phi}$ on both sides of 
\eq{NLEZFS} and  once more taking the  derivative with respect to $\hat{z}$ we reduce \eq{NLEZFS}
to the form:
\bea 
\frac{\chi(\gamma_{cr})}{1\,-\,\gamma_{cr}}\,\,\frac{d^2\,\phi}{(d z)^2}
\,\,\,&=&\,\,\,\left(\,1\,\,-\,\,e^{- \phi(z)}\,\right)\,\,-\,\,\frac{d \,L(\phi_z)}{d 
\,\phi_z}\,\frac{d^2\,\phi}{(d z)^2}\,\,;\label{NLEPHI1}\\
L(\phi_z)\,\,\,&=&\,\,\,\frac{\phi_z\,\,\chi\left(1\,\,-\,\,\phi_z 
\right)\,\,\,-\,\,\,1}{\phi_z}\,\,;\label{NLEPHI2}
\eea
The answer in the simple model given by \eq{SMK}  obtained in Ref. \cite{LT} corresponds to the 
simplified version of $\chi( 1 - f) 
= 1/f$. In this case the second term in \eq{NLEPHI1} vanishes and the solution for large $\hat{z}$ is 
\beq \label{NLEPHISOL}
\phi(\hat{z})\,\,=\,\,\frac{1\,-\,\gamma_{cr}}{\chi(\gamma_{cr})}\,\,\frac{z^2}{2}\,\,
\eeq

For the full BFKL kernel  function $ \frac{d \,L(\phi_z)}{d\,\phi_z} $ is negligibly small for small 
$\phi_z$. 
However, $\frac{d \,L(\phi_z)}{d\,\phi_z}$ being small for $\phi_z \,<\,1$ has a singularity at 
$\phi\,\,\rightarrow\,\,1$ namely
$$
\frac{d \,L(\phi_z)}{d\,\phi_z}\,\,\,\longrightarrow\,\,\,\,\frac{1}{\left(\,1\,\,-\,\,\phi_z 
\,\right)^2}\,\,\,\,\,\,\,\,\,\mbox{for}\,\,\,\,\,\,\,\,\,\,\,\phi_z \longrightarrow\,\,\,1
$$

The solution given by \eq{NLEPHISOL} is controversial since it was obtained assuming that $ \frac{d 
\,L(\phi_z)}{d\,\phi_z}$ is small but it  leads to large $\phi_z$  where this term is essential.
We can conclude that this solution is valid only in the simplified model with the kernel given by 
\eq{SMK}.

For the full BFKL kernel we can try an opposite approximation assuming that $\phi(z)$ is large 
($\phi\,\,\,\gg\,\,\,1$)  but 
$\phi_z$ is approaching to 1 at $z >>1$.
In vicinity $\phi_z = 1$ we reduce  \eq{NLEPHI1} to
\beq \label{EQREDU}
\frac{1}{\left( 1\,\,-\,\,\phi_z \,\right)^2}\,\,\,\frac{d^2\,\phi}{(d \hat{z})^2}\,\,=\,\,1.
\eeq

For large $z$ \eq{EQREDU} has a solution
\beq \label{NLEPHISOL1}
\phi(\hat{z})\,\,=\,\,z\,\,-\,\,\ln z
\eeq
which can be  verified  by explicit calculations. It should be stressed that $\phi(z)$ of 
\eq{NLEPHISOL1} satisfies all conditions of a smooth function that has been used for the derivation of 
\eq{NLEPHI1}. We can also check that the l.h.s. of \eq{NLEPHI1} is proportional to $1/z^2$ and it can 
be neglected.

It is more convenient to use the coordinate representation for the scattering amplitude ( see \eq{MR1} 
and 
\eq{MR2} ). We can simplify \eq{MR2} in the saturation region where  $N(r,y;b)$ manifests a geometrical 
scaling behaviour being a 
function of one variable $N(r^2Q^2_s(y,b))$. The main contribution in 
this equation stems from the kinematic region: $k\,r\,< 1$ and $r^2Q^2_s >1$. Indeed, for $k\,r\,>1$ 
the integral of \eq{MR2} is small,  due to the  oscillating behaviour of $J_0(k\,r)$ while for  
$r^2Q^2_s 
<1$
the amplitude $N$ being in the perturbative QCD domain is rather small. Therefore, \eq{MR2} can be 
rewritten in the form:
\beq \label{MR3}
\tilde{N}\left(k,y;b \right) \,\,=\,\,\tilde{N}\left( z 
\right)\,\,=\,\,\h\,\int^{\hat{z}}\,\,d\,z' N(z')
\eeq
where $z'$ in the integral is defined by \eq{ZR}. Comparing \eq{PHIMR},\eq{NLEPHISOL1} and  \eq{MR3} one 
concludes 
that
\beq \label{NRSOLFK}
N\left(\,r^2Q^2_s(y,b)\,\right)\,\,\,=\,\,\,\,1\,\,\,-\,\,\,e^{ - z(r)\,\,\,+\,\,\,\ln z(r)}
\eeq
with $z(r)$ is defined by 
\beq \label{ZSPA}
z(r)\,\,=\,\,\ln(Q^2_s\,r^2)\,\,=\,\,\bas \,\frac{\chi(\gamma_{cr})}{1
\,-\,\gamma_{cr}}\,\,(\,y\,\,-\,\,y_0\,)\,\,\,-\,\,\,\,\ln(1/r^2\Lambda^2)\,\,.
\eeq

\begin{figure}[htbp]
\begin{center}
\epsfig{file= 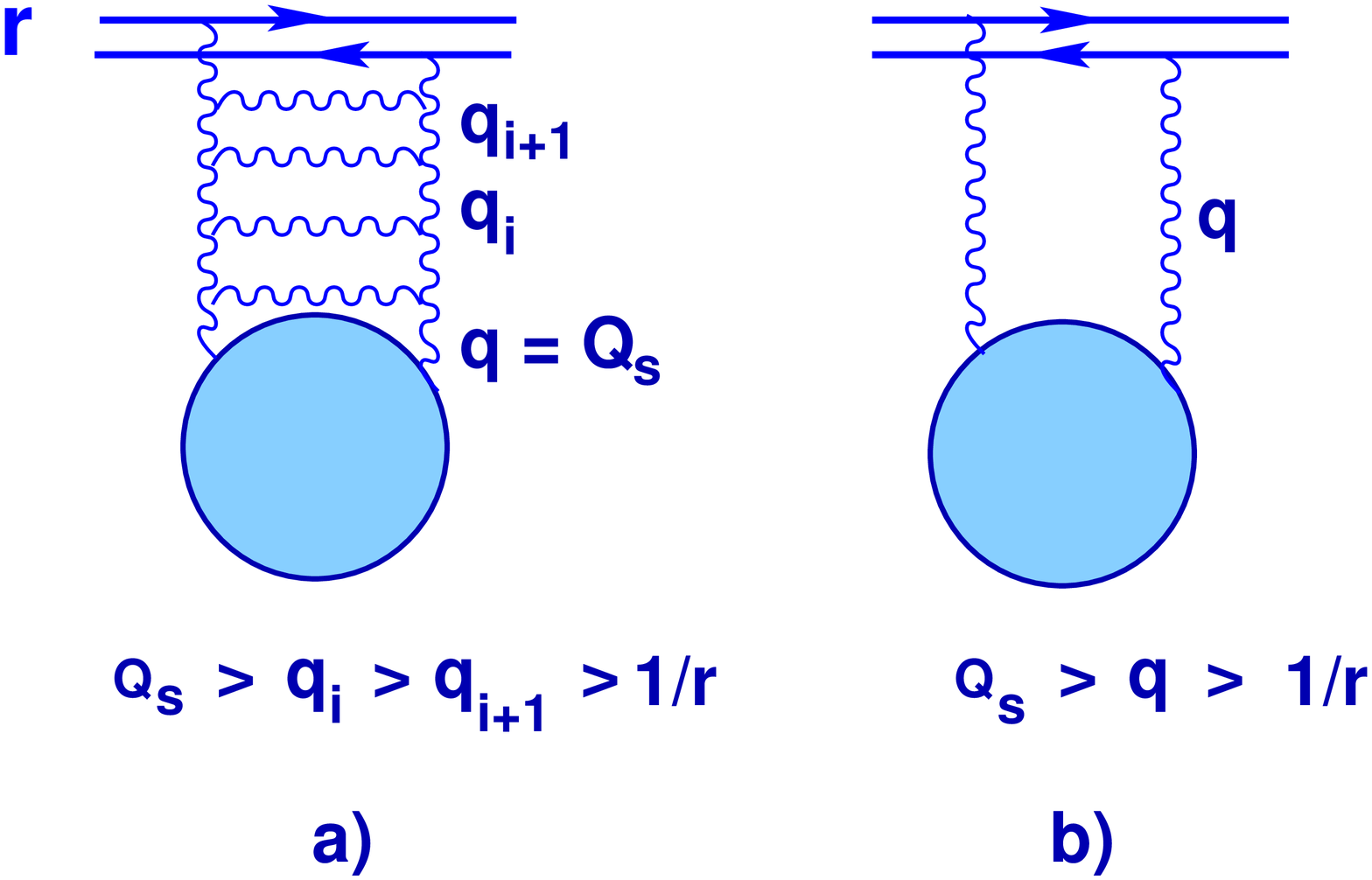,width=120mm}
\end{center}
\caption{The gluon emission which is suppressed in the solution given by \eq{NLEPHISOL} (\fig{basat}-a); 
and  the elastic rescattering (\fig{basat}-b) which is taken into account in the solution  by 
\eq{NRSOLFK} .}
\label{basat}
\end{figure}

We are aware that the majority of  experts, including one of us,  have thought that solution
of \eq{NLEPHISOL} gives a correct asymptotic behaviour for the scattering amplitude. Indeed, at first 
sight, the approach with the kernel of \eq{SMK} is well motivated \cite{LT}, while the same result can 
be derived without addressing any model for the kernel ( see Ref. \cite{IM} and references therein).
Function $\phi$ has a very transparent physical meaning (see \eq{SOL1}): $\exp^{- \phi}$ is a probability 
for a dipole to pass the target without any inelastic interaction. In the solution given by  
\eq{NLEPHISOL}, $\exp^{- \phi}$ is the probability that no extra gluon could be emitted (see 
\fig{basat}-a). However,  this solution does not suppress the elastic scattering shown in 
\fig{basat}-b. This diagrams can easily  be calculated
\beq \label{BAS}
N_{el}\,\,\propto\,\,\frac{1}{\pi\,r^2}\,\,\int^{\frac{1}{r^2}}_{Q^2_s}\,\,\frac{d 
q^2}{q^4}\,\left(\frac{q^2}{Q^2_s} \right) \,\,=\,\,e^{ - z(r)\,\,\,+\,\,\,\ln z(r)}
\eeq
where  $q^2/Q^2_s$ describes the emission of the `soft' gluon with $1/r < q< Q_s $ from the dipole with 
the typical size $1/Q_s$. Since $N_{el}$ is the amplitude at fixed impact parameter we divided the cross 
section by the area of the scattering dipole. The idea that approaching to the black disc will be 
described by the Born approximation in the saturation region has been discussed \cite{LERY} but we found 
that this asymptotic behaviour follows from the Balitsky-Kovchegov non-linear equation.

\begin{figure}[htb]
\begin{minipage}{10cm}
\begin{center}
\epsfig{file= 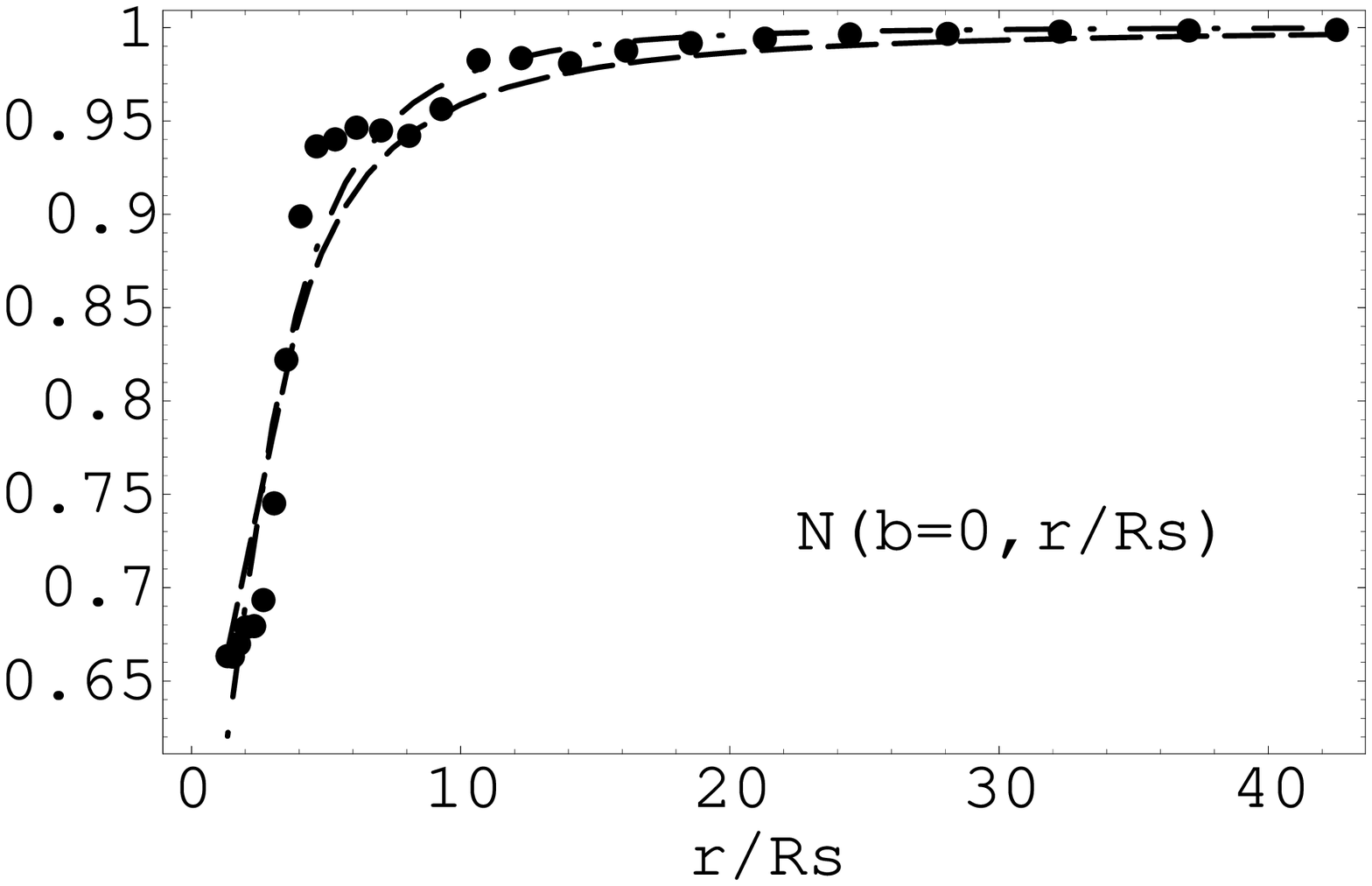,width=90mm}
\end{center}
\end{minipage}
\begin{minipage}{6cm}
\caption{The fit of the numerical solution to the Balitsky-Kovchegov equation \cite{GLMSOL} using 
\eq{FIT1} and \eq{FIT2}. The dashed line corresponds to \eq{FIT2} with
 $C^2_1 = 0.69$ and $C^2_2 = -1.05$ 
while the dashed-dotted line describes \eq{FIT1} with $C^1_1 = 0.25$ and 
$C^1_2 = 0.11$ .}
\label{numsol}
\end{minipage}
\end{figure}

A kind of argument against could be the  result of the  numerical simulations for
dipole-dipole scattering  by Salam in Ref. \cite{MS} which supports  \eq{NLEPHISOL},  in spite of the
fact that
the general BFKL kernel has been used. In \fig{numsol} we compare two solutions (\eq{NLEPHISOL} and 
\eq{NRSOLFK} ) with the numerical solution of the Balitsky-Kovchegov equation given in Ref.\cite{GLMSOL}.
We fit the numerical solutions by two function: 
\bea \label{FIT}
N(z)\,\,&=&\,\,1 \,\,-\,\,C^1_1\,e^{ - C^1_2 z^2}\,\,;\label{FIT1}\\
N(z)\,\,&=&\,\,1 \,\,-\,\,C^2_1\,e^{ - (z - ln(z 
+C^2_2))}\,\,;\label{FIT2}
\eea
One can see that both formulae could describe the solution. Therefore, we cannot use the numerical 
solution as an argument in favour of any solution. It is interesting to mention that the value of $C_2$ 
in
\eq{FIT1} ($C_2 = 0.11$) is closer to the theoretical value in \eq{NLEPHISOL} than it is  found by 
Salam.

\subsubsection{High energy amplitude in the saturation region:}

~

\begin{boldmath}
{\bf  Fixed $\as$ :}
\end{boldmath}

~

We can calculate the scattering amplitude in the saturation region using the Iancu - Mueller 
factorization in the form of \eq{IMFORFAL} but with the contour $C_1$ for integration over $\nu$.
For $N(Y - y,r,b; [\nu]) $ we  use \eq{NRSOLFK} with 
$ z$ replaced by $ z \,\,+\,\,2 \Phi[\gamma(r_i)]$ where $\Phi$ is   defined 
in \eq{PHIGA} and  $z\, =\, \ln \,Q^2_s\,\,\,-\,\,\,\ln(1/r^2)$.  As in the case of the double 
log kernel we choose $\gamma(r_i) = \nu$ and $ \gamma(r'_i) \,=\,\tilde{\gamma}(r_i,r'_i)/\nu$.

We can easily take the integral of \eq{IMFORFAL}  if we neglect the $\nu$ dependence 
in $\ln z$. In 
this case the product $N(Y - y,r,b[\nu] \,N(y,R,b;[\tilde{\gamma}(r_i,r'_i)/\nu])$  does not depend on 
$\nu$. This simple observation leads to the answer for the scattering amplitude
\beq \label{SAFK}
N\left(\,r, R,Y;b\,\right)\,\,=\,\,1\,\,\,-\,\,\,e^{-\,z(Y,r,R) + 2 \,\ln \left(z(Y,r,R)/2 \right)}
\eeq 
where $z(Y,r,R)$ is given by \eq{ZRR}.

As in the case of the  simplified kernel of \eq{SMK} ,  
 the main contribution stems from the set of enhanced diagrams.

~

\begin{boldmath}
{\bf Running $\as$ :}
\end{boldmath}

~

Calculating the diagram of \fig{basat}-b for the running QCD coupling we obtain the answer 
\beq \label{AMR}
\phi (y,r) \,\,=\,\,\ln(r^2\,Q^2_s) \,\,-\,\,\ln \left(\ln \left( 
\frac{\ln(Q^2_s/\Lambda)}{\ln(1/(r^2\,\Lambda^2))}\right) \right)
\eeq
with $Q_s \propto e^{\sqrt{cy}}$. We can also verify that \eq{AMR} is the solution to the 
generalization of the equation of 
\eq{NLEQMR} with the prescription for running $\as$ given in Ref. \cite{MU02}.

This  change leads to  remarkable
consequences.  Indeed, since the minimal value of $Z(Y - y, r) \,+\,Z( y,R)$, 
where $Z(y,R)\,\,\ln(Q^2_s(y)\,R^2)$, 
occurs at $y=0$ or 
$Y-y = 0$ the resulting amplitude in the saturation region is
\beq \label{SAFKRUN}
N\left(\,r, R,Y;b\,\right)\,\,=\,\,1\,\,\,-\,\,\,e^{-\,\ln(Q^2_s(Y)/\Lambda^2) + \ln(R^2/r^2) }
\eeq
For simplicity,  we did not write in \eq{SAFKRUN} the $\ln \ln (\dots$ in \eq{SAFKRUN}.  

Therefore, the asymptotic behaviour of the scattering amplitude for running $\as$ is determined by the 
non-linear equation \cite{B,K},  which is able to  describe  the colour glass condensate phase of QCD 
\cite{CGC} in this case.

\section{Conclusions}
In this paper we showed that the Iancu-Mueller factorization \cite{IM} is closely related to a sum of  a 
particular 
set of enhanced diagrams (see \fig{gefunenh}) in the Reggeon -like diagram technique  based on the BFKL 
Pomeron. This set of diagrams leads to the high energy scattering amplitude in wide range of energy 
and/or Bjorken $x$, namely
\beq \label{RAPCON}
Y = \ln(1/x) \,\,<\,\,\frac{1}{\bas}\,\ln\,\frac{N^2_c}{\bas}\,\,.
\eeq
 
We found a simple formula of \eq{IMFORFAL} which sums these diagrams and manifests the 
Iancu-Mueller factorization. Using this formula we confirm the Iancu-Mueller result, that
at fixed QCD coupling, the high energy asymptotic behaviour is determined by the enhanced diagrams, but 
not by the `fan' diagrams that are  summed by the means of Balitsky-Kovchegov non-linear equation. We 
showed that the dipole -dipole amplitude in the saturation region can be written as
\beq \label{NFIN}
N(r,R;Y)\,\,=\,\,1 - \left(\,N^{BK}(r,R;Y) - 1\,\right)^{\frac{1}{2}}\,\,.
 \eeq
However, for the running QCD coupling  the main contribution in the saturation region 
stems from the `fan' diagrams  of \fig{fandi}-a (or \fig{gefunfan}) -type.  This new result allows us
to penetrate the saturation region,  summing the `fan' diagrams of more general type (see \fig{fandi}-b 
for the first such diagram). The first suggestion of how to sum such diagrams was proposed in Ref. 
\cite{LL}, and the present   paper confirms our hope that the general `fan' diagrams will lead to the 
estimates of 
high energy asymptotic behaviour  for running QCD coupling   in the saturation region.

Our final result for the asymptotic behaviour of the scattering amplitude is 
quite different from that in the Iancu - Mueller paper \cite{IM} even for fixed QCD coupling case. This 
difference 
stems from the new solution to the non-linear equation  found in this paper,  which includes the general 
BFKL kernel as opposed  to the simplified version of the kernel ( see \eq{SMK} ) which has 
been used in 
Ref. \cite{LT}.

Our approach to the Iancu-Mueller factorization is even closer to the 
Monte Carlo simulation than the original papers \cite{IM}, and the semi-analytical estimates suggested 
in Ref. \cite{MS}. We hope that the 
simple formula suggested in this paper will give an impetus for searching for a new approach to find  
the  high 
energy 
amplitude in the saturation region. We believe that we have demonstrated in this paper, that the 
Iancu-Mueller factorization provides  a method  to take  the essential fluctuations in the 
partonic 
wave function of colliding dipoles  into account and shows the practical method of calculation beyond of 
the 
non-linear equation. It also allows us to find  the region of applicability of the non-linear 
equation.

\section*{Acknowledgments}

  We thank   E. Gotsman, U. Maor and A. Mueller for fruitful and stimulating 
discussions on the subject of this paper. We are  specially grateful to A. Mueller that he drew out 
attention to his ideas on how to include the QCD  running  coupling constant 	in the non-linear equation 
\cite{MU02}. 

  E.L.  is indebted to the Alexander-von-Humboldt Foundation for the
award that gave him a possibility to work on low $x$ physics during the
last year.

 This research was supported in part  by the Israel Science Foundation,
founded by the Israeli Academy of Science and Humanities.

\end{document}